\documentclass[prd,aps,onecolumn]{revtex4-1}
\usepackage{amsfonts}
\usepackage{amsmath}
\usepackage{graphicx}
\usepackage{subfig}

\begin{document}
\title{First stage of LISA data processing II: Alternative filtering dynamic models for LISA}
\author{Yan Wang}
\email{yan.a.wang@uwa.edu.au}
\affiliation{The University of Western Australia, School of Physics, 35 Stirling Highway,
Crawley, Perth, Western Australia 6009, Australia}
\affiliation{previously at Max-Planck-Institut f\"ur Gravitationsphysik (Albert-Einstein-Institut), Callinstra{\ss}e 38, 30167 Hannover, Germany}

\author{Gerhard Heinzel}
\email{gerhard.heinzel@aei.mpg.de}
\affiliation{Max-Planck-Institut f\"ur Gravitationsphysik (Albert-Einstein-Institut), Callinstra{\ss}e 38, 30167 Hannover, Germany}

\author{Karsten Danzmann}
\email{karsten.danzmann@aei.mpg.de}
\affiliation{Max-Planck-Institut f\"ur Gravitationsphysik (Albert-Einstein-Institut), Callinstra{\ss}e 38, 30167 Hannover, Germany}

\begin{abstract}
Space-borne gravitational wave detectors, such as (e)LISA, are designed to operate in the
low-frequency band (mHz to Hz), where there is a variety of gravitational wave sources
of great scientific value \cite{Danzmann13,Babak11}. To achieve the extraordinary sensitivity of these detector,
the precise synchronization of the clocks on the separate spacecraft and the
accurate determination of the interspacecraft distances are important ingredients.
In our previous paper \cite{Wang14a}, we have described a hybrid-extend Kalman filter
with a full state vector to do this job. In this paper, we explore several
different state vectors and their corresponding (phenomenological) dynamic models,
to reduce the redundancy in the full state vector, to accelerate the algorithm,
and to make the algorithm easily extendable to more complicated scenarios.
\end{abstract}
\maketitle

\section{Introduction}

Unlike ground-based gravitational wave (GW) detectors \cite{LIGO92,ligo,VIRGO97}, space-borne GW detectors
\cite{LISA98,Danzmann03,LISA11,Danzmann13,Kawamura06,Ando2010,Crowder,Wang13,Gong11,Ni13} usually have varying
interspacecraft distance and the clock based in each spacecraft (S/C) [as the vertex of the detector] are freely running individually. To achieve the extraordinary target sensitivities of the detectors, precise determination of interspacecraft distances and accurate synchronization of the separate spacecraft clocks in different S/C are usually necessary. This happens in the first stage of data processing on ground.

We have tried to solve this for (e)LISA \cite{LISA98,Danzmann03,LISA11,Danzmann13} in the preprocessing stage in our previous paper \cite{Wang14a}, where a hybrid-extended Kalman filter was designed to preprocess raw interspacecraft measurements of (e)LISA. Although this ranging determination and clock synchronization problem shares some similarity with the global positioning system (GPS) track problem \cite{GPS,Mohinder08}, the differences \cite{Wang14a} between them make it far more complicated than the GPS case, see \cite{Wang14a} for detailed discussions.
The 24-dimensional state vector designed in \cite{Wang14a} cannot be determined well due to the insufficient number of measurements. In this paper, we will explore alternative dynamic models for Kalman filters to reduce the ambiguity in the state vector, while using the same measurements. We investigate in detail how different filtering dynamic models affect the accuracy of the estimates. These various dynamic models also provide better means to deal with other effects, such as jittering recording time, or the Sagnac differential delay.

In Sec. \ref{S:meas}, we describe the inter-spacecraft measurements that are going to be used to improve the ranging
accuracy and to synchronize the clocks in different S/C. We also review the full 24-dimensional physical state vector used for Kalman filters in paper \cite{Wang14a}. In Sec. \ref{S:K23} and \ref{S:K22}, we try to model the system with 23-dimensional and 22-dimensional state vectors by excluding the common absolute time errors and the frequency errors of the clocks. In Sec. \ref{S:KS}, we describe a toy system model with only the clock variables. Although this toy model is too simplified for realistic simulation, it allows us to see the analytic forms of all the components of the measurement matrix $H$ and the dynamic matrix $F$. It may provide insights to more complicated problem in reality.
In Sec. \ref{S:sin} and \ref{S:poly}, we explore two effective models with phenomenological parameters to further
reduce the ambiguity in the state vector. The effective models try to use variables that are directly measured,
thus reducing the nonlinearities in the measurement equations and increasing the numerical
accuracies of the Kalman filter algorithms. The number of variables in the state vector
of these effective models is also significantly smaller, and their matrices
are better conditioned. It turns out that the effective models are simpler and more efficient.
Finally, we summarize the results in Sec. \ref{S:sum}.

\section{Interspacecraft measurements and the full state vector}
\label{S:meas}

The interspacecraft measurements \cite{Wang14a,Barke10,Gerhard11} of (e)LISA contain information about the absolute distance between
the S/C, and their clock errors. This information is expected to be extracted from the measurements in the pre-processing stage \cite{Wang14a} of (e)LISA data analysis. At S/C $j$, the local carrier laser with frequency $f^\textrm{carrier}_j$
beats with the incoming laser from S/C $i$. The beatnotes $D_{ij}$ contain information of the Doppler shifts, hence we call them the Doppler measurements,
which are given as follows
\begin{eqnarray}\label{eq:D}
D_{ij} &=& \left[  f^\textrm{carrier}_j - f^\textrm{carrier}_i \left( 1 - \frac{(\vec{v}_j-\vec{v}_i)\cdot \hat{n}_{ij}}{c} \right) \right. \nonumber \\
&+& \left. f^\textrm{GW}_{ij}\right]\left( 1- \frac{\delta f_j}{f^\textrm{nom}_j}\right)+n^D_{ij},
\end{eqnarray}
where $\vec{v}_i$ is the velocity of S/C $i$, $c$ is the speed of light, $f^\textrm{carrier}_i$ denotes the carrier frequency of the laser from S/C $i$,
$\hat{n}_{ij}$ denotes the unit vector along the axis from S/C $i$ to $j$, $f^\textrm{nom}_i$ and $\delta f_i$ are the nominal frequency and the frequency error of the clock, and $n^D_{ij}$ is the measurement noise. Since this measurement contains the expected gravitational wave signals $f^\textrm{GW}_{ij}$, it is also referred
to as the science measurement.

Pseudo-random codes are modulated onto the carrier lasers to measure the distance between the S/C \cite{Esteban11}.
By correlating the remote laser with an electronic copy of the same pseudo-random code in the phasemeter, the propagation time
of the laser is determined, similar to the pseudorange in the global navigation satellite system \cite{GPS,Mohinder08}. We call these the ranging measurements and denote them as follows
\begin{eqnarray}\label{eq:R}
R_{ij} = L_{ij} + (\delta T_j - \delta T_i) \cdot c + n^R_{ij},
\end{eqnarray}
where $L_{ij}$ denotes the arm lengths, and $\delta T_i$ denotes the time error of the clock in S/C $i$, which is
directly related to the frequency error \cite{Wang14a}.

The clock frequencies are up-converted to the GHz range and also modulated onto the carrier lasers. The sidebands thus created are called
the clock sidebands. They contain the information of the frequency errors of the clocks. They measure
the same frequency change as the carrier-to-carrier beatnotes, and, in addition, they measure the amplified differential clock errors. Here, we
use the simplified formulae of the clock measurements obtained after subtracting the common-mode component $D_{ij}$ and dividing by the
up-conversion factor \cite{Wang14a}
\begin{eqnarray}\label{eq:C}
C_{ij} = \delta f_j - \delta f_i + n^C_{ij}.
\end{eqnarray}
More detailed descriptions of the interspacecraft measurements can be found in \cite{Wang14a} and the references therein.

By writing the subscripts explicitly, we have 18 measurements in total
\begin{eqnarray}
y &=& (R_{31},D_{31},C_{31},R_{21},D_{21},C_{21},R_{12},D_{12},C_{12}, \nonumber \\
      &&R_{32},D_{32},C_{32},R_{23},D_{23},C_{23},R_{13},D_{13},C_{13})^T.  \nonumber
\end{eqnarray}
However, to fully describe the system in this aspect by physical variables,
we need a 24-dimensional column state vector \cite{Wang14a}
\begin{eqnarray}
x=(\vec{x}_1, \vec{x}_2, \vec{x}_3, \vec{v}_1, \vec{v}_2, \vec{v}_3, \delta T_1, \delta T_2, \delta T_3,
\delta f_1, \delta f_2, \delta f_3)^T,   \nonumber
\end{eqnarray}
where $\vec{x}_i=(x_i,y_i,z_i)^T$ are the positions of the S/C in the solar system, and $\vec{v}_i=(v_{xi},v_{yi},v_{zi})^T$ their velocities.
It is clear at this point that we have more unknown variables than measurements in the Kalman filter model. This is obvious from the fact that the
relative measurements between the S/C do not provide information on the absolute positions and velocities of the S/C.

\section{A Kalman filter model with a 23 dimensional state vector}
\label{S:K23}

In particular, the clock errors $\delta T_i$ appear in the measurement equations only in the form
of differential time errors $\delta T_j - \delta T_i$. Therefore, the commom mode of the clock errors $(\delta T_1 + \delta T_2 + \delta T_3 )/3$ cannot be determined. To eliminate this degeneracy in the Kalman filter model, we replace the three clock errors $\delta T_1$, $\delta T_2$ and $\delta T_3$ by
two differential clock errors $\delta T_1 - \delta T_2$ and $\delta T_2 - \delta T_3$. In consequence, the state vector is now a
23-dimensional column vector
\begin{eqnarray}
x&=&(\vec{x}_1, \vec{x}_2, \vec{x}_3, \vec{v}_1, \vec{v}_2, \vec{v}_3, \delta T_1 - \delta T_2, \delta T_2 - \delta T_3, \nonumber \\
&& \delta f_1, \delta f_2, \delta f_3)^T. \nonumber
\end{eqnarray}
The third differential clock-error variable $\delta T_1 - \delta T_3$ can be expressed as $ (\delta T_1 - \delta T_2) + ( \delta T_2 - \delta T_3) $.
The dynamic equation for the clock errors can simply be modified as
\begin{eqnarray}
\frac{\mathrm{d}}{\mathrm{d}t}(\delta T_j - \delta T_i) = \frac{\delta f_j}{f_j^\mathrm{nom}} - \frac{\delta f_i}{f_i^\mathrm{nom}}.
\end{eqnarray}
The dynamic matrix $F_k$ and the observation matrix $H_k$ are then modified accordingly.

We carried out simulations to compare the performance of this Kalman filter model with the performance of the model designed in paper \cite{Wang14a} with
a 24-dimensional state vector. The two filters have been run over the same simulated measurement data. Fig.~\ref{fig:L23} shows a comparison of the arm length determination between the two Kalman filter models. It can be seen that the performance of the two Kalman filter models is comparable in determining the arm lengths. Both models can successfully decouple the arm lengths from the initial clock biases and reduce the noise in the arm-length roughly by one order of magnitude.
Notice that the performance of a Kalman filter depends on the specific noise realization. Therefore, a small difference between
the two models in the estimation error is insignificant. Fig.~\ref{fig:T23} shows histograms of estimation errors in differential clock errors.
Both models are able to reduce the differential clock errors roughly by one order of magnitude. The Kalman filter model with a 23-dimensional state vector performs slightly better. Fig.~\ref{fig:f23} shows histograms of estimation errors in differential frequency errors. Both Kalman filter models have greatly reduced the noise in the raw measurements. The performances turn out to be comparable.

All in all, the Kalman filter model with a 23-dimensional state vector designed in this section performs slightly better than the model with a
24-dimensional state vector. The reason is that the Kalman filter model with a smaller state vector has reduced the ambiguity in the system model, while
retaining the full information on the measurement mechanism.

\begin{widetext}

\begin{figure}[htbp]
\centering
\subfloat[]{
\begin{minipage}[t]{0.4\textwidth}
\centering
\includegraphics[width=1.0\textwidth]{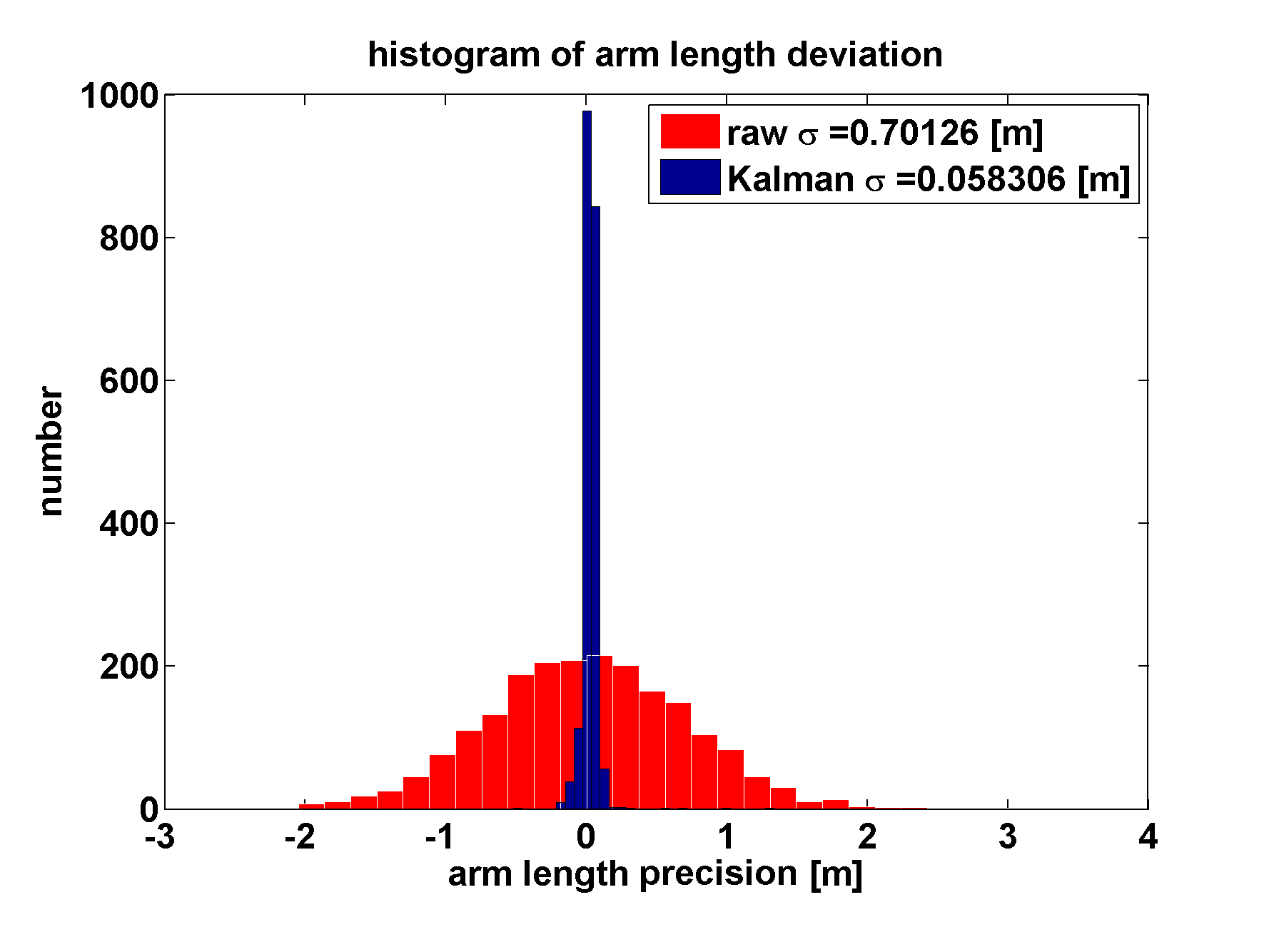}
\end{minipage}
}
\subfloat[]{
\begin{minipage}[t]{0.4\textwidth}
\centering
\includegraphics[width=1.0\textwidth]{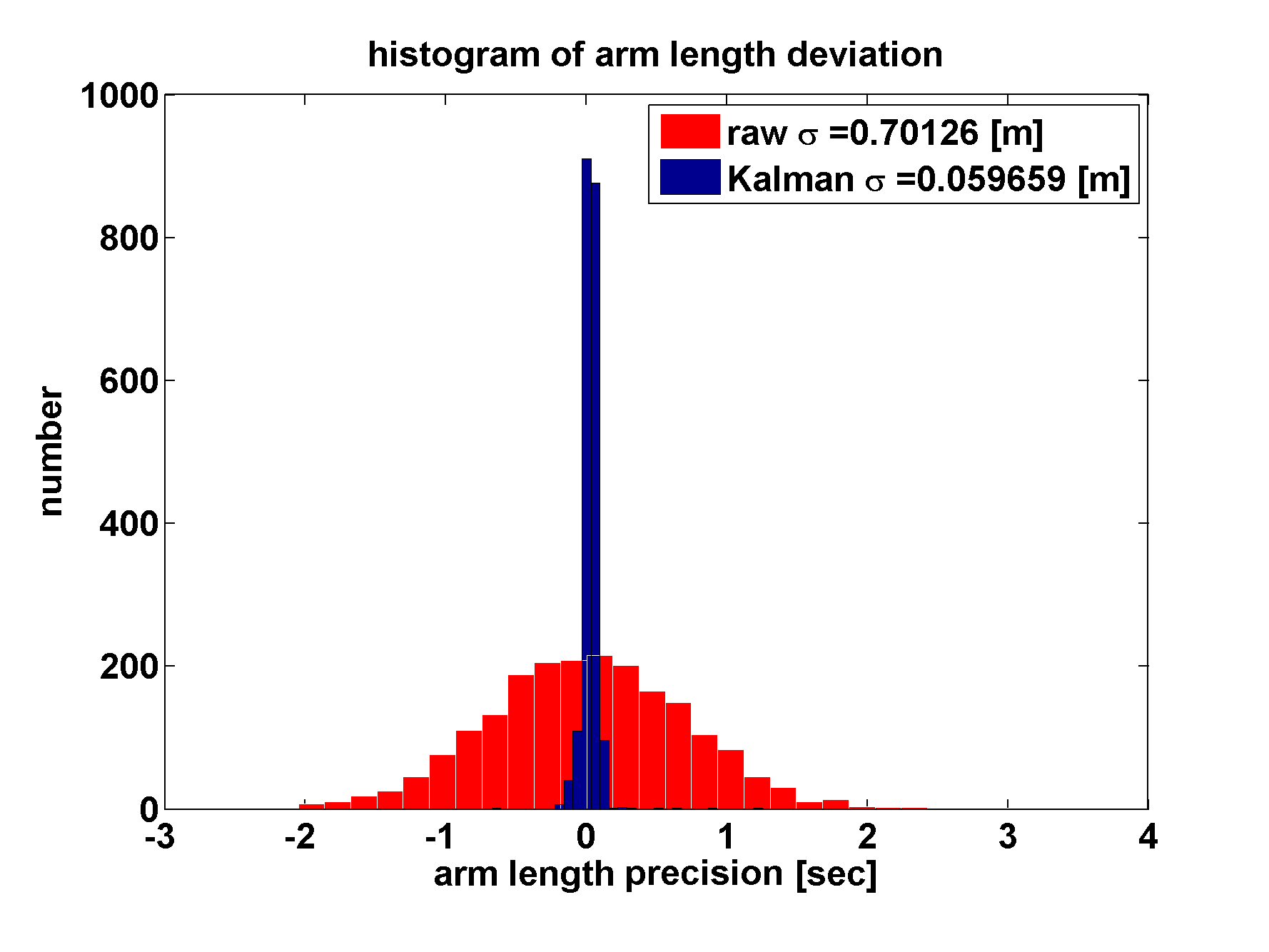}
\end{minipage}
}
\caption{ \label{fig:L23} A comparison of the arm length determination. Fig.~(a) shows histograms of errors in raw arm-length measurements
and Kalman filter estimates with a 24-dimensional state vector. Fig.~(b) shows histograms of errors in raw arm-length measurements
and Kalman filter estimates with a 23-dimensional state vector. Notice that the initial clock biases are not included in the raw measurement
errors for better vision.
}
\end{figure}

\begin{figure}[htbp]
\centering
\subfloat[]{
\begin{minipage}[t]{0.4\textwidth}
\centering
\includegraphics[width=1.0\textwidth]{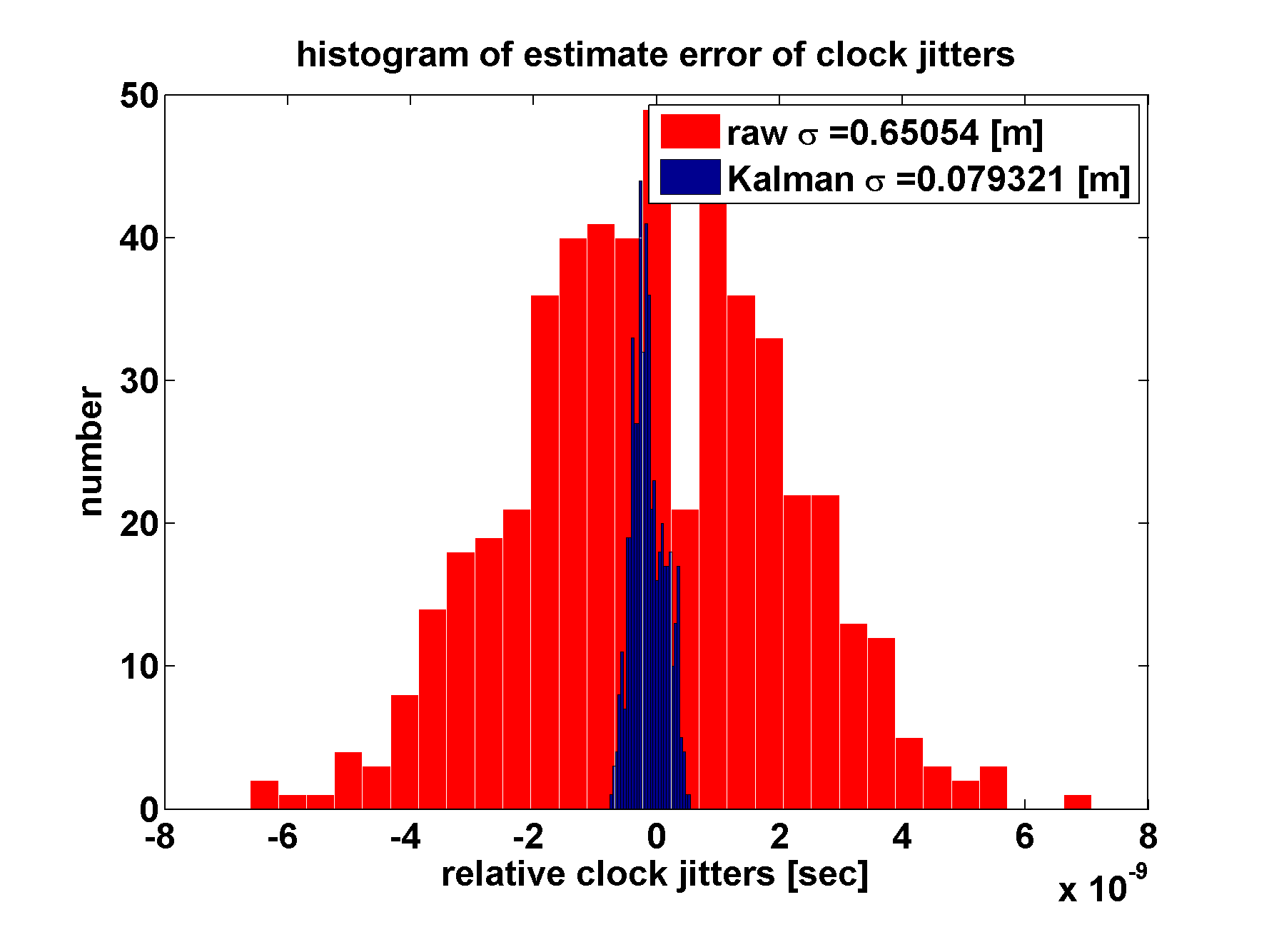}
\end{minipage}
}
\subfloat[]{
\begin{minipage}[t]{0.4\textwidth}
\centering
\includegraphics[width=1.0\textwidth]{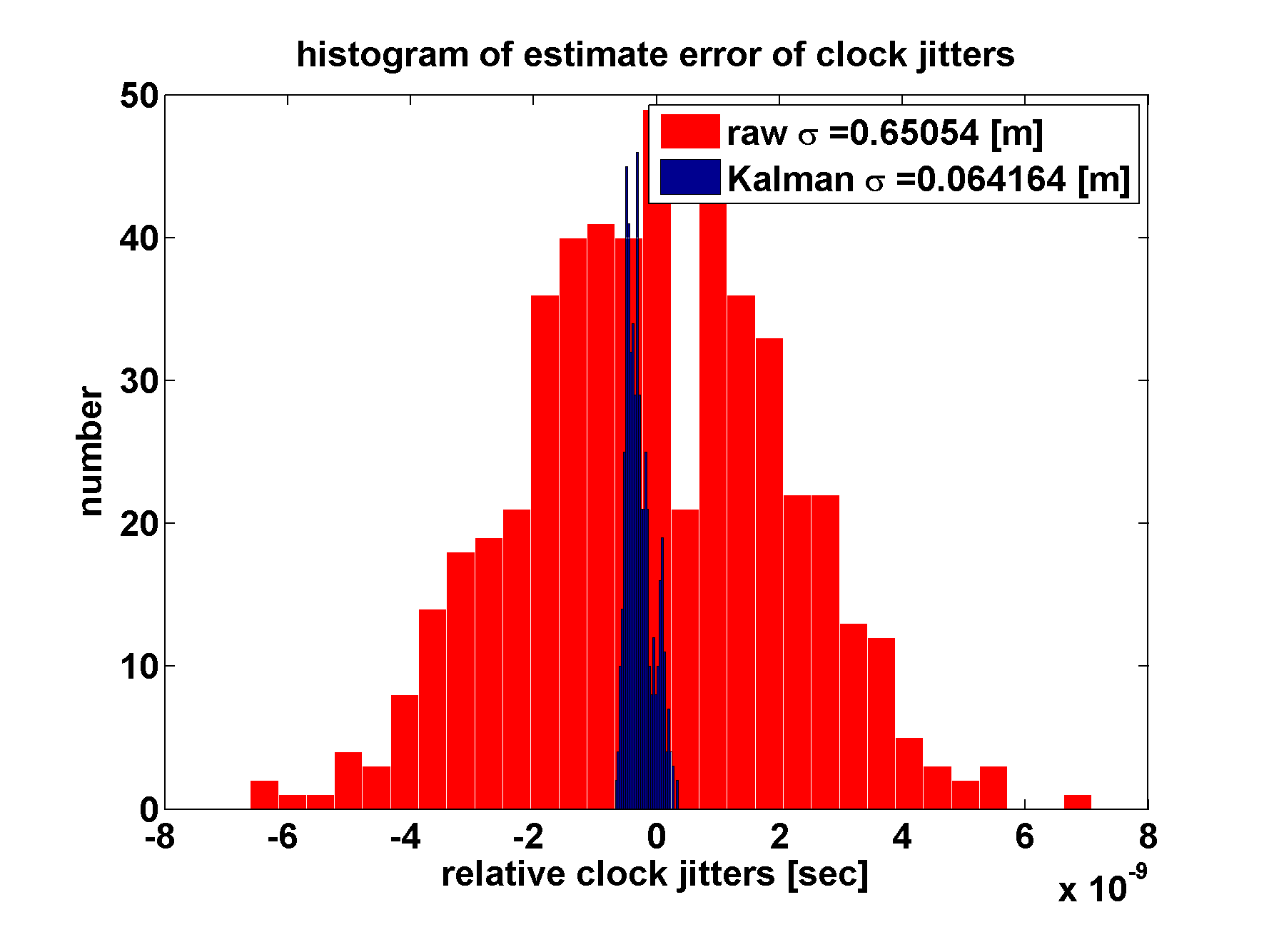}
\end{minipage}
}
\caption{ \label{fig:T23} Histograms of estimation errors in differential clock errors. Fig.~(a) shows histograms of errors in the raw data
and Kalman filter estimates with a 24-dimensional state vector. Fig.~(b) shows histograms of errors in the raw data and Kalman filter
estimates with a 23-dimensional state vector.
}
\end{figure}

\begin{figure}[htbp]
\centering
\subfloat[]{
\begin{minipage}[t]{0.4\textwidth}
\centering
\includegraphics[width=1.0\textwidth]{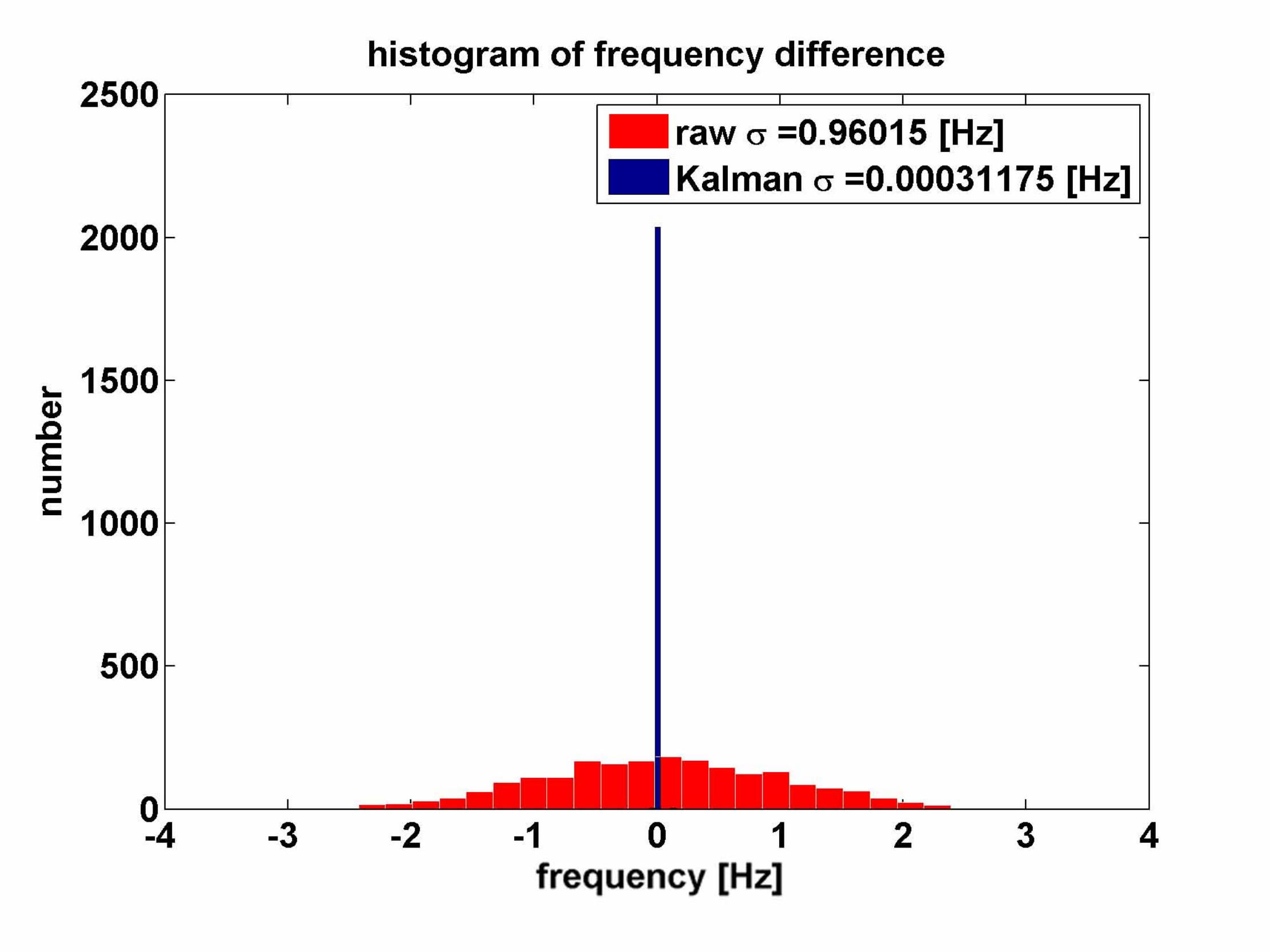}
\end{minipage}
}
\subfloat[]{
\begin{minipage}[t]{0.4\textwidth}
\centering
\includegraphics[width=1.0\textwidth]{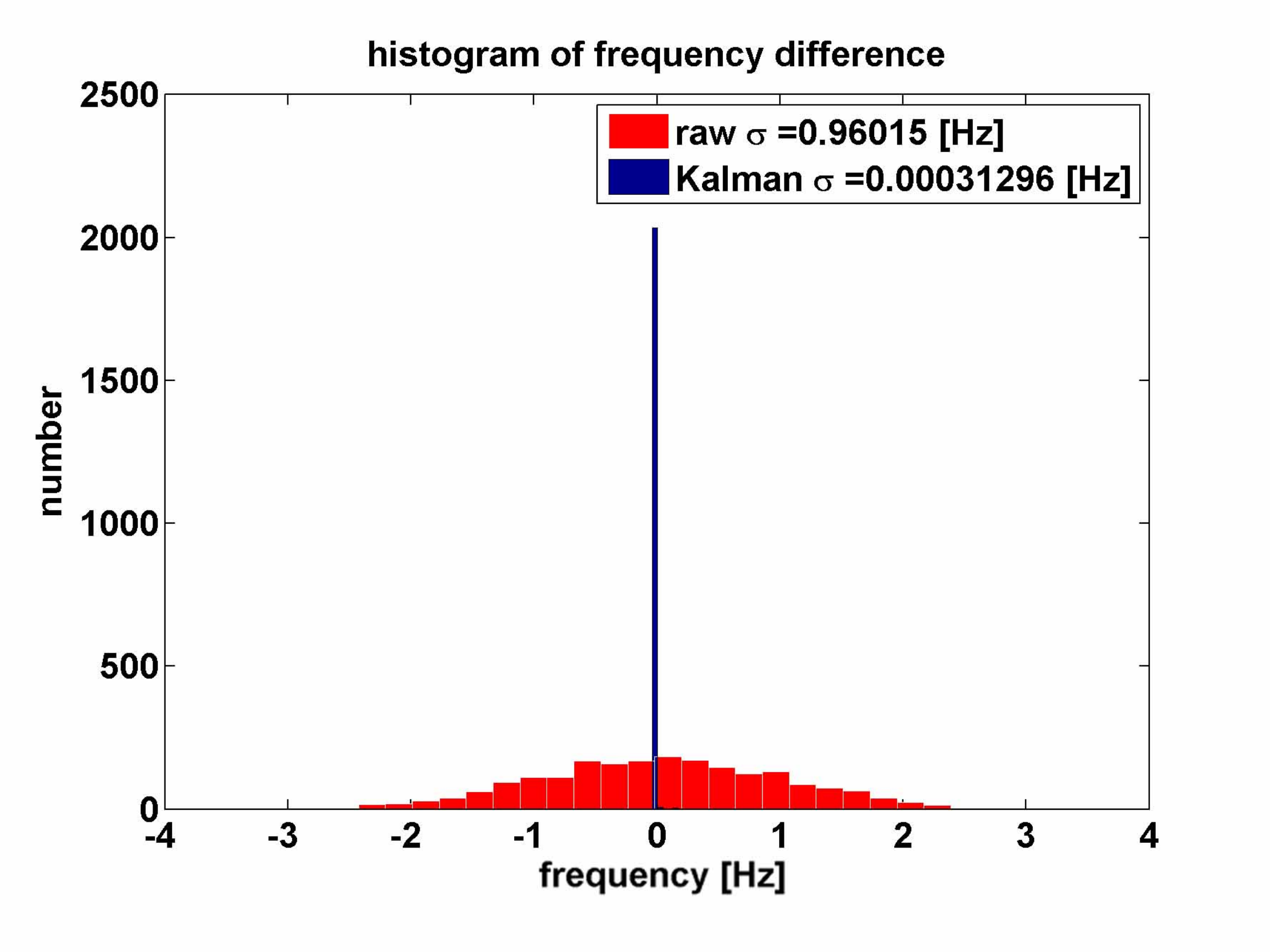}
\end{minipage}
}
\caption{ \label{fig:f23} Histograms of estimation errors in differential frequency errors of the clocks. Fig.~(a) shows histograms of errors in the raw data
and Kalman filter estimates with a 24-dimensional state vector. Fig.~(b) shows histograms of errors in the raw data and Kalman filter
estimates with a 23-dimensional state vector.
}
\end{figure}

\begin{figure}[htbp]
\centering
\subfloat[]{
\begin{minipage}[t]{0.33\textwidth}
\centering
\includegraphics[width=1.0\textwidth]{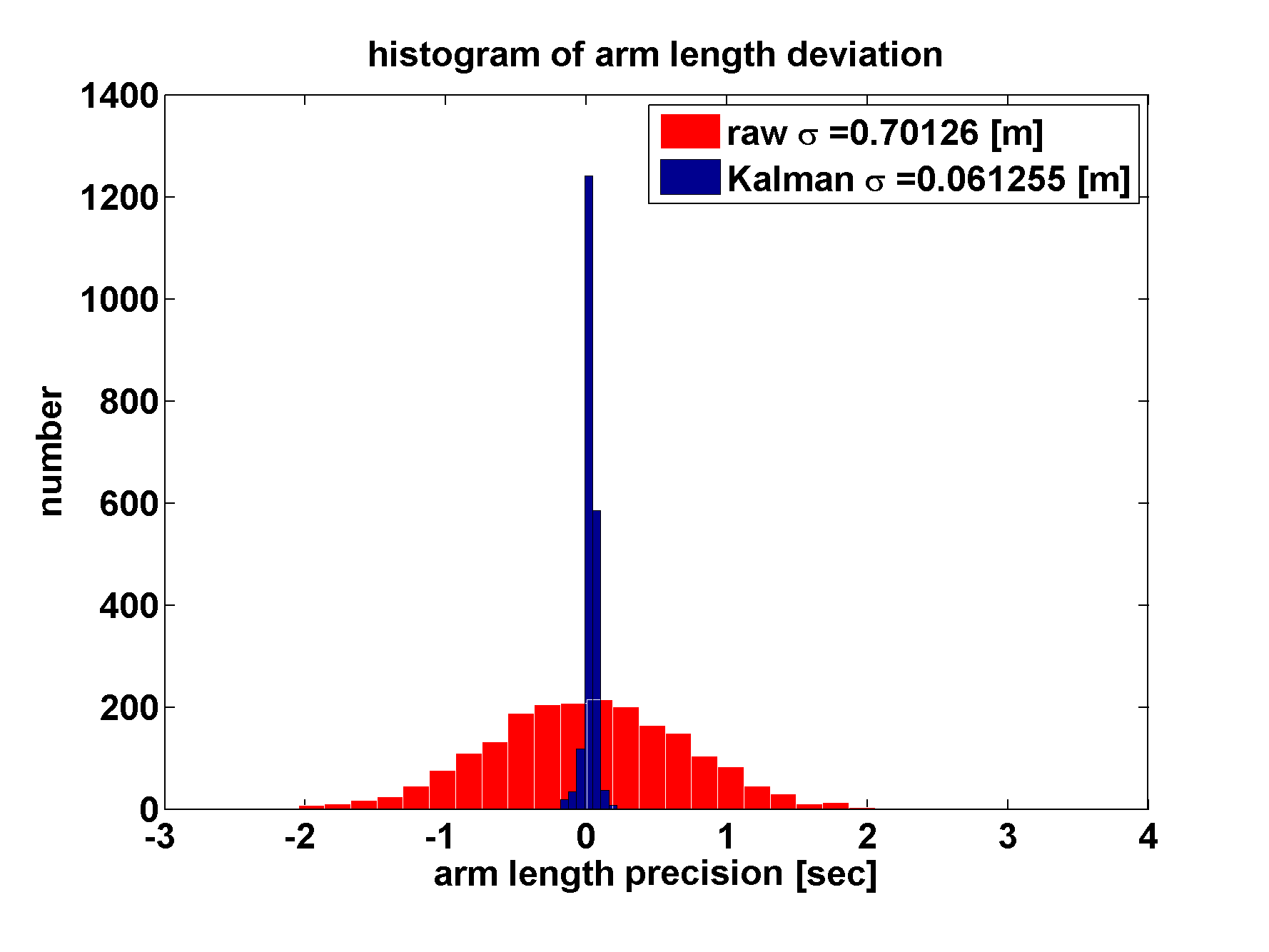}
\end{minipage}
}
\subfloat[]{
\begin{minipage}[t]{0.33\textwidth}
\centering
\includegraphics[width=1.0\textwidth]{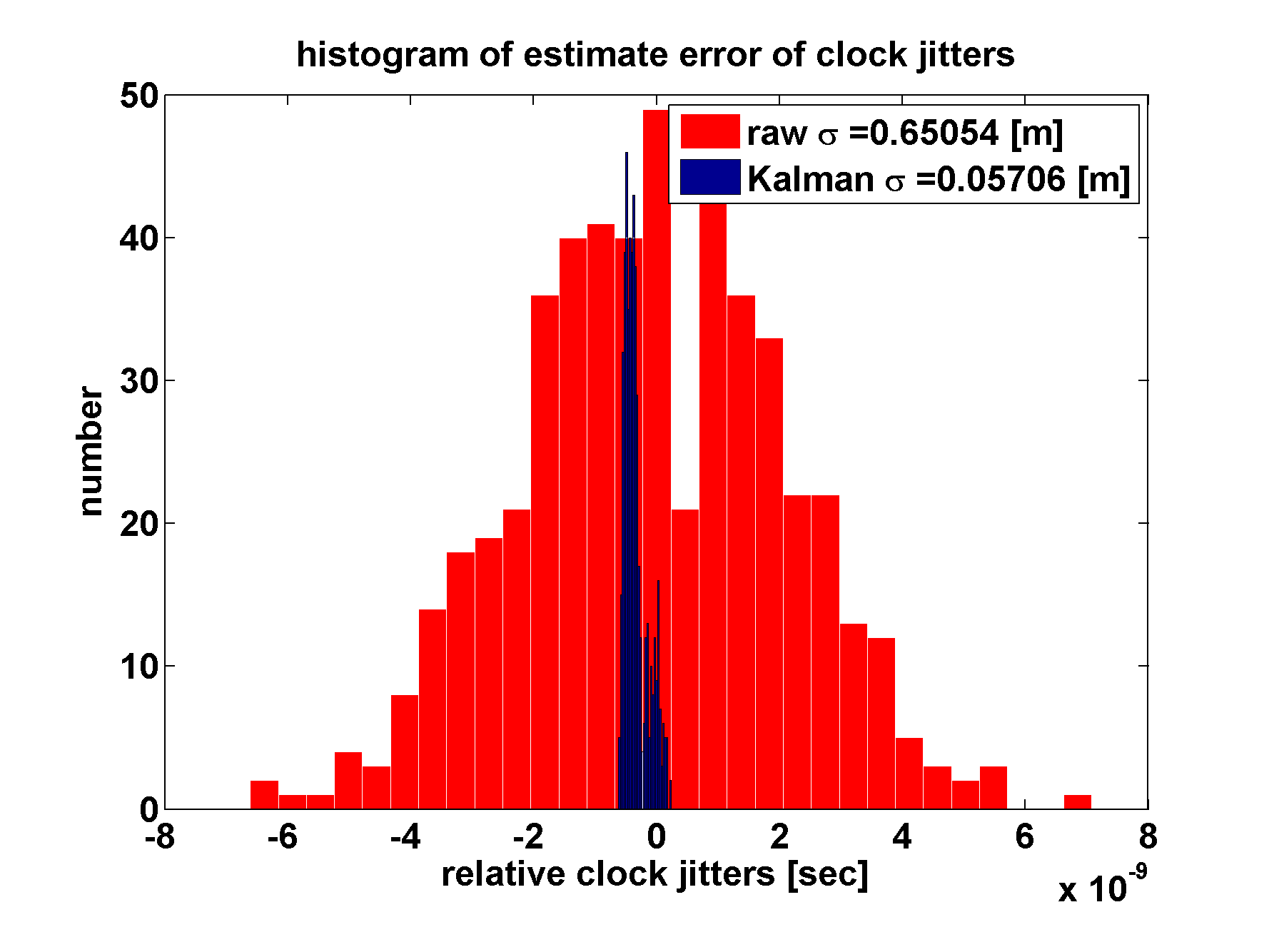}
\end{minipage}
}
\subfloat[]{
\begin{minipage}[t]{0.33\textwidth}
\centering
\includegraphics[width=1.0\textwidth]{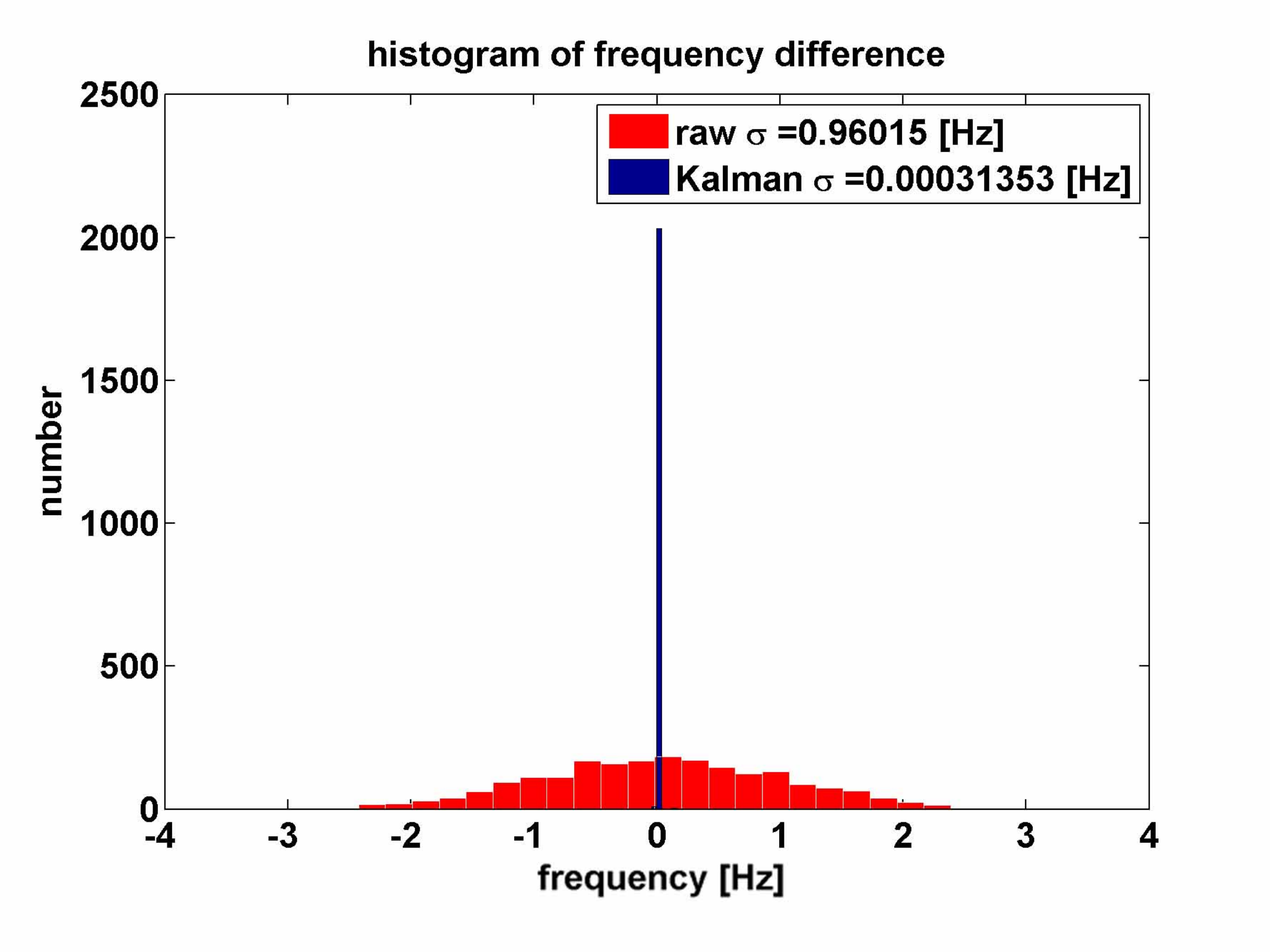}
\end{minipage}
}
\caption{ \label{fig:Kalman22} Simulation results of the Kalman filter model with a 22-dimensional state vector. Fig.~(a) histograms of errors in raw arm-length measurements and Kalman filter estimates. Fig.~(b) shows histograms of the raw and residual differential clock errors.  Fig.~(c) shows histograms of the raw and residual differential frequency errors.}
\end{figure}

\end{widetext}

\section{A Kalman filter model with a 22 dimensional state vector}
\label{S:K22}

In this section, we try to further simplify the state vector. From the simulation, we know that the absolute frequency error $\delta f_i$ cannot
be determined accurately, since the common mode of the three frequency-error variables is poorly determined by the measurements. (Here, the `absolute' frequency error is in contrast with the differential frequency error $\delta f_i - \delta f_j$.) The only information about the absolute frequency errors comes from the term $1 - \delta f_j/f^\textrm{nom}_j$ of Eq. \ref{eq:D}.
Usually, $\delta f_j/f^\textrm{nom}_j$ is several orders of magnitude smaller than 1, thus this term provides very limited information.

We can actually approximate this term by 1 without losing much information. Then, the Doppler measurement equation becomes
\begin{eqnarray}\label{eq:Dsimp}
D_{ij} &\doteq& \left[  f^\textrm{carrier}_j - f^\textrm{carrier}_i \left( 1 - \frac{(\vec{v}_j-\vec{v}_i)\cdot \hat{n}_{ij}}{c} \right) + f^\textrm{GW}_{ij}\right] \nonumber \\
&& + n^D_{ij}.
\end{eqnarray}
Now, in all the three interspacecraft measurement equations~\ref{eq:R}, \ref{eq:Dsimp} and \ref{eq:C} the frequency errors only appear
in the form of differential frequency errors $\delta f_j - \delta f_i$. Therefore, we can replace the three frequency error variables $\delta f_i$
by two differential frequency error variables $\delta f_1 - \delta f_2$ and $\delta f_2 - \delta f_3$. The state vector is correspondingly reduced to 22 dimensions:
\begin{eqnarray}
x&=&(\vec{x}_1, \vec{x}_2, \vec{x}_3, \vec{v}_1, \vec{v}_2, \vec{v}_3, \delta T_1 - \delta T_2, \delta T_2 - \delta T_3, \nonumber \\
&& \delta f_1 - \delta f_2, \delta f_2 - \delta f_3)^T. \nonumber
\end{eqnarray}
The dynamic model for the clock errors and the frequency errors is approximated as follows
\begin{eqnarray}
\frac{\mathrm{d}}{\mathrm{d}t}
\begin{bmatrix}
\delta T_j - \delta T_i \\
\delta f_j - \delta f_i
\end{bmatrix}
=
\begin{bmatrix}
2 (\delta f_j - \delta f_i) / ( f_j^\mathrm{nom} + f_i^\mathrm{nom} ) \\
0
\end{bmatrix}.
\end{eqnarray}
The main advantages of this Kalman filter model are (i) the further reduction of the near degeneracy in the system model,
(ii) the reduction of the nonlinearity in the measurement equations. On the other hand, the use of approximate
measurement equations could also be disadvantageous at the same time.

We implemented this Kalman filter model to process the same simulated measurement data used in the last section.
We have set the nominal frequencies of the clocks in different S/C to be identical in the simulation, i.e. $f_j^\mathrm{nom} = f_i^\mathrm{nom}$.
If the nominal frequencies are not identical, the performance of this Kalman filter model would be degraded depending on how different the
nominal frequencies are. The simulation results are summarized in Fig.~\ref{fig:Kalman22}, where the estimation errors in the arm lengths, clock errors and their frequency errors are shown. Comparing these results with Fig.~\ref{fig:L23}, \ref{fig:T23} and \ref{fig:f23}, we find that the estimation errors of this model in the arm lengths are slight larger than those of the other two Kalman filter models. However, this Kalman filter model performs slightly better than the other two
models in determining the clock errors. Overall, the performances are similar.


\section{A toy Kalman filter model with only clock variables}
\label{S:KS}

In this section, we show a toy Kalman filter model only for the clocks. It is simple and can be implemented efficiently. However, this model is accurate only
when the interspacecraft laser links are instantaneous, hence not straightforward to be generalized to the complete treatment
with relativistic interspacecraft delays. At any rate, in this model we need not to deal with large matrices with hundreds of nonlinear
components but only small constant matrices. The benefits of studying such a toy model are the following: (i) the matrices in this
model are of much smaller size and can be shown explicitly with clear dependence on the variables, hence it helps the audience to understand
how Kalman filters work for (e)LISA; (ii) with the clock variables separated from the arm-length and the Doppler measurements, it can be used to design and tune sophisticated clock models more efficiently.

We first define a 4-dimensional state vector
\begin{eqnarray}
x =
\begin{bmatrix}
\delta T_1 - \delta T_2 \\
\delta T_2 - \delta T_3 \\
\delta f_1 - \delta f_2 \\
\delta f_2 - \delta f_3
\end{bmatrix}
\equiv
\begin{bmatrix}
x_1 \\
x_2 \\
x_3 \\
x_4
\end{bmatrix},
\end{eqnarray}
which consists of only the differential clock errors and the differential frequency errors. The dynamic equations can be
simply modeled as
\begin{eqnarray}
\frac{\mathrm{d}}{\mathrm{d}t}
\begin{bmatrix}
x_1 \\
x_2 \\
x_3 \\
x_4
\end{bmatrix}
=
\begin{bmatrix}
x_3/f^\mathrm{nom} \\
x_4/f^\mathrm{nom} \\
0   \\
0
\end{bmatrix}
+ \mathrm{noise},
\end{eqnarray}
where we have assumed the nominal frequencies of the three clocks are identical, i.e. $f^\mathrm{nom}_i=f^\mathrm{nom}$.
Since these dynamic equations are linear, the dynamic matrix is simply
\begin{eqnarray}
F = \frac{\partial f}{\partial x} =
\begin{bmatrix}
0 & 0 & 1/f^\mathrm{nom} & 0 \\
0 & 0 & 0 & 1/f^\mathrm{nom} \\
0 & 0 & 0 & 0 \\
0 & 0 & 0 & 0
\end{bmatrix}.
\end{eqnarray}

Next, we set up the measurement equations. For instantaneous laser links, the differential clock errors
can be obtained in the following way
\begin{eqnarray}
\delta T_j - \delta T_i = \frac{1}{2c}(R_{ij} - R_{ji} - n^R_{ij} + n^R_{ji} ),
\end{eqnarray}
where the ranging measurements $R_{ij}$ are given in Eq.~\ref{eq:R}. The differential frequency errors of the clocks are
indirectly inferred by the modified clock measurements, see Eq.~\ref{eq:C}. As mentioned before, the Doppler
measurements (see Eq.~\ref{eq:D}) contain little information on the frequency errors, hence for this Kalman
filter model with only clock variables they are almost irrelevant. Therefore, the measurement equations are
simplified to the following linear form
\begin{eqnarray}
y =
\begin{bmatrix}
\delta T_1 - \delta T_2 \\
\delta T_2 - \delta T_3 \\
\delta T_3 - \delta T_1 \\
\delta f_1 - \delta f_2 \\
\delta f_2 - \delta f_3 \\
\delta f_3 - \delta f_1
\end{bmatrix}
+ \mathrm{noise}
=
\begin{bmatrix}
x_1 \\
x_2 \\
- x_1 - x_2   \\
x_3 \\
x_4 \\
- x_3 - x_4
\end{bmatrix}
+ \mathrm{noise},
\end{eqnarray}
which leads to a constant measurement matrix
\begin{eqnarray}
H =  \frac{\partial y}{\partial x} =
\begin{bmatrix}
1 & 0 & 0 & 0 \\
0 & 1 & 0 & 0 \\
-1& -1& 0 & 0 \\
0 & 0 & 1 & 0 \\
0 & 0 & 0 & 1 \\
0 & 0 & -1&-1
\end{bmatrix}.
\end{eqnarray}

We implement the above simplified Kalman filter model to process the same simulated measurement data used in previous sections.
The simulation results are shown in Fig.~\ref{fig:Ksimp}. It can be seen that in the considered case the simplified Kalman filter
has significantly reduced the noise in the differential clock errors and greatly reduced the noise in the differential frequency errors.
Comparing these results with Fig.~\ref{fig:T23}, \ref{fig:f23} and \ref{fig:Kalman22}, the performance of this simplified Kalman filter
is a few times worse than other Kalman filter models in determining the relative clock jitter and about ten times worse in determining the
differential frequency errors of the clocks.


\section{A periodic system model}
\label{S:sin}

If each S/C of LISA would follow a Kepler orbit, the variations of the arm lengths of LISA were strictly periodic. Even for
relativistic orbits with gravity contributions from all the planets in the solar system, the variations of the arm lengths are also mostly periodic. Using the Kepler orbital setup in the paper \cite{Dhurandhar05}, the annual evolutions of the LISA arm lengths  are shown in Fig.~\ref{fig:armlength} (a).
The arm-length variations resemble sinusoidal functions within 6 months. In the subsequent
half year, the arm length variation remains a roughly sinusoidal shape, but with a different amplitude.
By optimizing the inclination angle between the ecliptic plane and the LISA constellation plane \cite{Nayak06}, the arm-length
variations can be further reduced, and we obtain the arm-length variations shown in Fig.~\ref{fig:armlength} (b).
These arm-length variations are similar to a sinusoidal function with a period of one year.
In either cases, we can phenomenologically model the arm-length evolutions as follows
\begin{eqnarray}
L(t) = \bar{L} + A \sin(\omega t),
\end{eqnarray}
where $\bar{L}$ is the average arm length, and $\omega$ is angular frequency with either a half-year period
or a one-year period. If we define the arm length change as below
\begin{eqnarray}
\Delta L(t) \equiv L(t) - \bar{L} = A \sin(\omega t),
\end{eqnarray}
the dynamic equation of the arm length change is the same as that of a simple harmonic oscillation
\begin{eqnarray}\label{eq:HarmEq}
\frac{\mathrm{d}^2 \Delta L}{\mathrm{d}t^2} + \omega^2 \Delta L = 0.
\end{eqnarray}

\begin{widetext}

\begin{figure}[htbp]
\centering
\subfloat[]{
\begin{minipage}[t]{0.4\textwidth}
\centering
\includegraphics[width=1.0\textwidth]{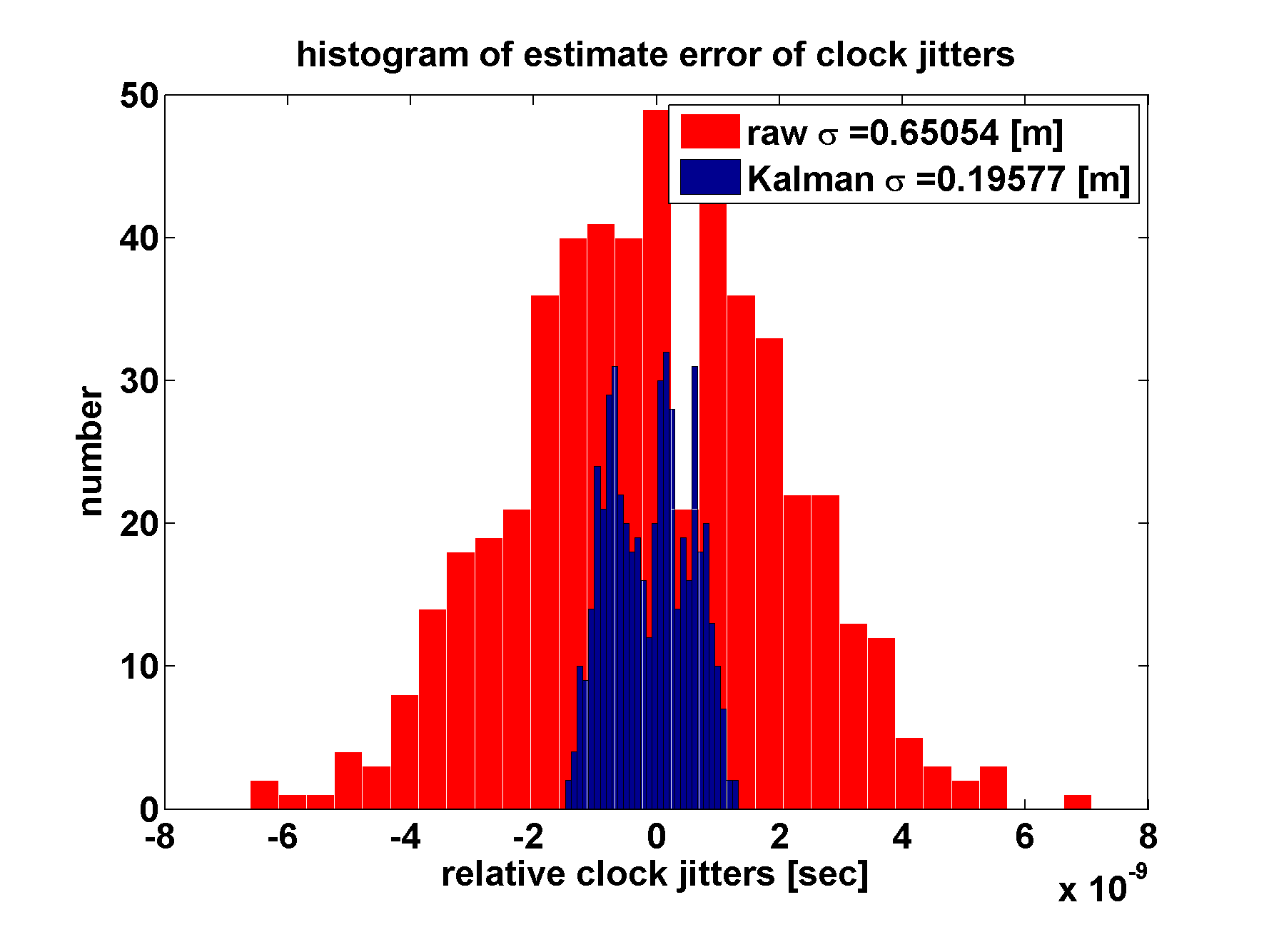}
\end{minipage}
}
\subfloat[]{
\begin{minipage}[t]{0.4\textwidth}
\centering
\includegraphics[width=1.0\textwidth]{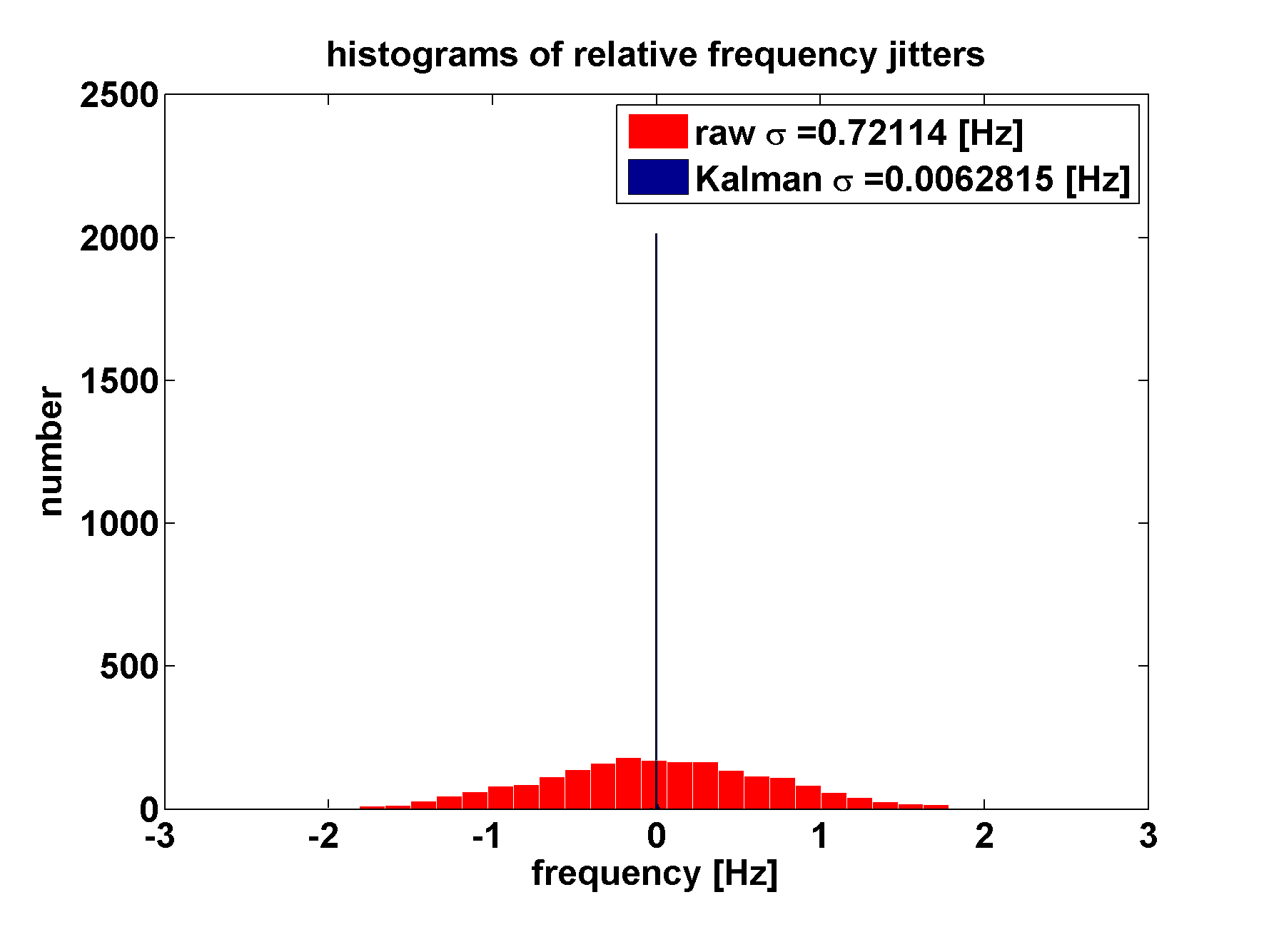}
\end{minipage}
}
\caption{ \label{fig:Ksimp} Simulation results of the simplified Kalman filter model with only clock variables. Fig.~(a) shows histograms of errors in the
differential clock errors.  Fig.~(b) shows histograms of errors in the differential frequency errors of the clocks.}
\end{figure}

\begin{figure}[htbp]
\centering
\subfloat[]{
\begin{minipage}[t]{0.4\textwidth}
\centering
\includegraphics[width=1.0\textwidth]{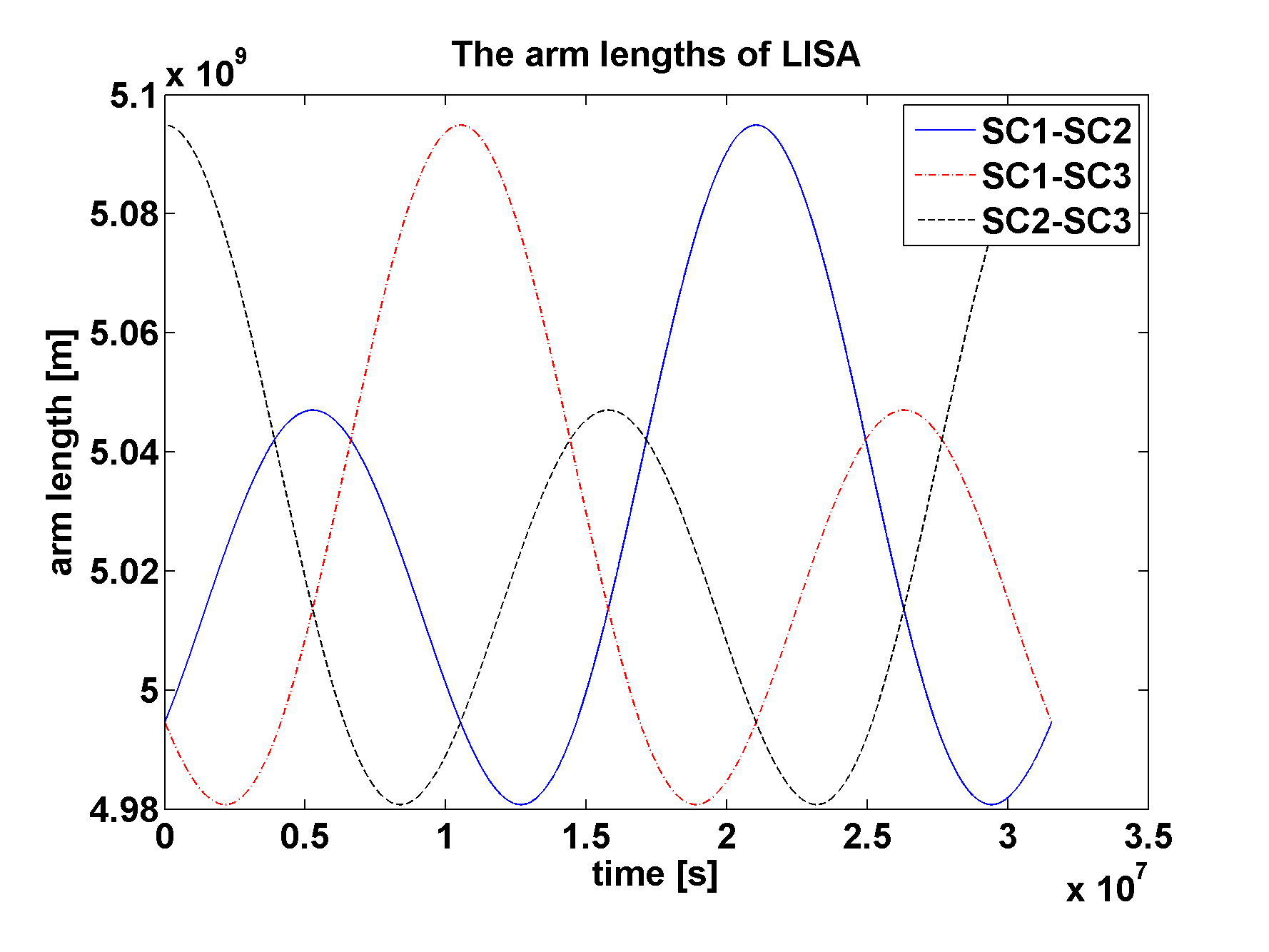}
\end{minipage}
}
\subfloat[]{
\begin{minipage}[t]{0.4\textwidth}
\centering
\includegraphics[width=1.0\textwidth]{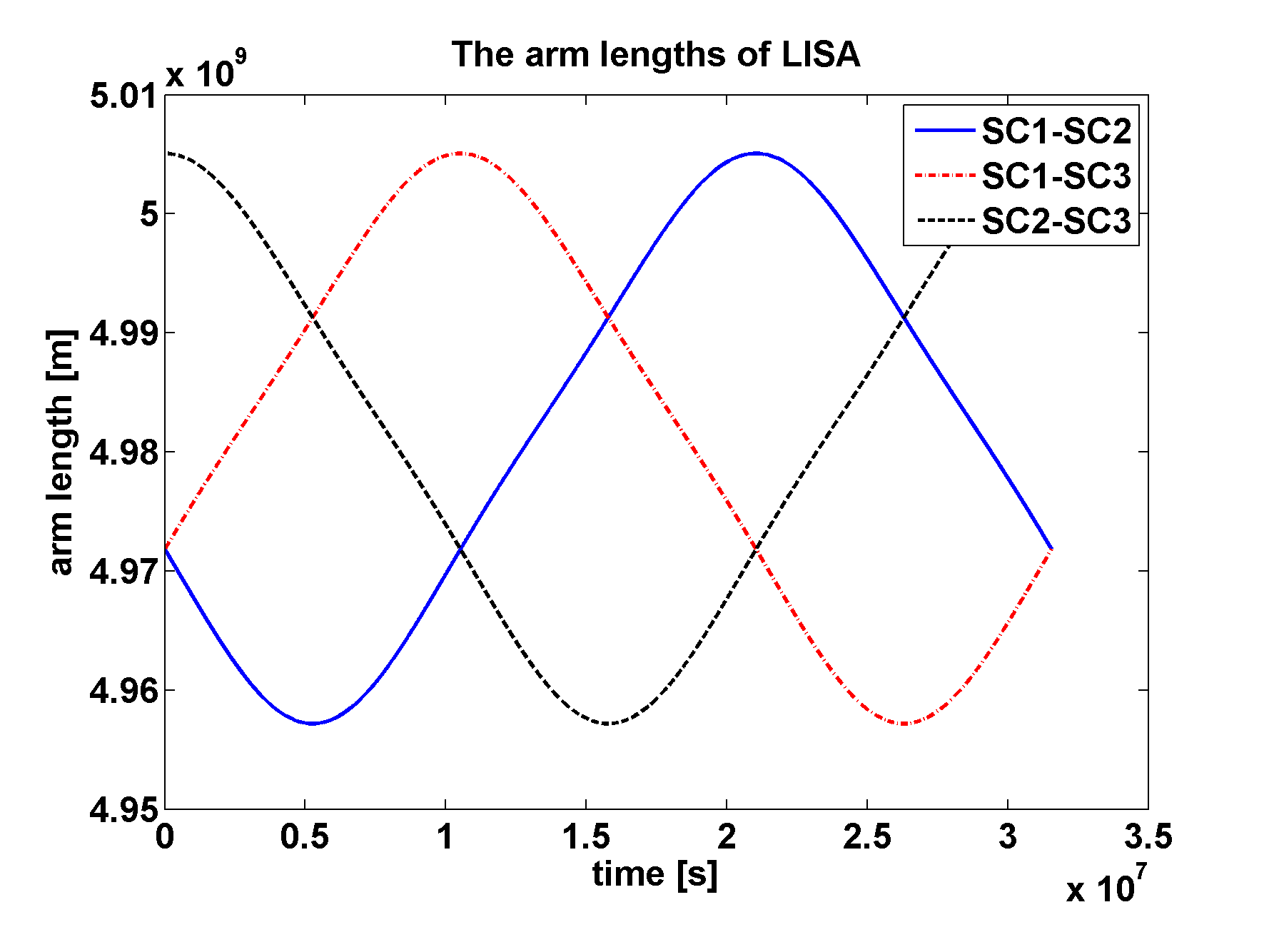}
\end{minipage}
}
\caption{ \label{fig:armlength} The annual arm-length variations of the LISA constellation for Kepler orbits. Fig.~(a) has fixed the
inclination angle between the ecliptic plane and the constellation plane of LISA as $60^\circ$. Fig.~(b) has varied and optimized this inclination
angle to minimize the annual arm-length variations.}
\end{figure}

\end{widetext}

A Kalman filter based on this simple periodic dynamic model can thus be designed. We define a 11-dimensional
state vector as follows
\begin{eqnarray}
x&=&(\Delta L_{21}, v_{21}, \Delta L_{32}, v_{32}, \Delta L_{13}, v_{13}, \nonumber \\
&& \delta T_1 - \delta T_2,\delta T_2 - \delta T_3, \delta f_1, \delta f_2, \delta f_3)^T. \nonumber
\end{eqnarray}
The dynamic equations for the arm-length variations and the relative tangential velocities can be obtained from Eq.~\ref{eq:HarmEq}
by rewriting it as first order differential equations
\begin{eqnarray}
\frac{\mathrm{d}}{\mathrm{d}t}
\begin{bmatrix}
\Delta L \\
v
\end{bmatrix}
=
\begin{bmatrix}
0 & 1 \\
-\omega^2 & 0 \\
\end{bmatrix}
\begin{bmatrix}
\Delta L \\
v
\end{bmatrix}
+ \mathrm{noise}.
\end{eqnarray}
The dynamic equations for the clock variables are given as follows
\begin{widetext}
\begin{eqnarray}\label{eq:ClockEq}
\frac{\mathrm{d}}{\mathrm{d}t}
\begin{bmatrix}
\delta T_1 - \delta T_2 \\
\delta T_2 - \delta T_3 \\
\delta f_1 \\
\delta f_2 \\
\delta f_3
\end{bmatrix}
=
\begin{bmatrix}
0 & 0 & 1/f_1^\mathrm{nom} & -1/f_2^\mathrm{nom} & 0 \\
0 & 0 & 0 & 1/f_2^\mathrm{nom} & -1/f_3^\mathrm{nom} \\
0 & 0 & 0 & 0 & 0 \\
0 & 0 & 0 & 0 & 0 \\
0 & 0 & 0 & 0 & 0
\end{bmatrix}
\begin{bmatrix}
\delta T_1 - \delta T_2 \\
\delta T_2 - \delta T_3 \\
\delta f_1 \\
\delta f_2 \\
\delta f_3
\end{bmatrix}
+ \mathrm{noise}. \nonumber \\
\end{eqnarray}
\end{widetext}

In order to make the system model fit better into the Kalman filter frame, we subtract the average arm length from
the ranging measurements, thus obtaining the following observation vector
\begin{eqnarray}
y' &=& (R_{31} - \bar{L},D_{31},C_{31},R_{21} - \bar{L},D_{21},C_{21},R_{12} - \bar{L}, \nonumber \\
&& D_{12},C_{12},R_{32} - \bar{L},D_{32},C_{32},R_{23} - \bar{L}, \nonumber \\
&& D_{23},C_{23},R_{13} - \bar{L},D_{13},C_{13})^T, \nonumber
\end{eqnarray}
where the interspacecraft measurements are
\begin{eqnarray}
R_{ij} &=& L_{ij} + (\delta T_j - \delta T_i) c + n^R_{ij}, \\
D_{ij} &=& \left[  f^\textrm{carrier}_j - f^\textrm{carrier}_i \left( 1 - \frac{v_{ij}}{c} \right) + f^\textrm{GW}_{ij}\right]\left( 1- \frac{\delta f_j}{f^\textrm{nom}_j}\right)  \nonumber \\
&& + n^D_{ij}, \\
C_{ij} &=& \delta f_j - \delta f_i + n^C_{ij}.
\end{eqnarray}
The Measurement matrix $H'_k = \left.\frac{\partial h'_k}{\partial x}\right|_{\hat{x}_k^-}$ is $18$-by-$11$. We explicitly
give its first three rows below
\begin{widetext}
\begin{eqnarray}
\begin{bmatrix}
0 \dots 0 & 1 & 0 & c & c & 0 & 0 & 0 \\
0 \dots 0 & 0 & \frac{f_3^\textrm{carrier}}{c}\left( 1- \frac{\delta f_1}{f^\textrm{nom}_1}\right) & 0 & 0 & \frac{f^\textrm{carrier}_3-f^\textrm{carrier}_1}{f^\textrm{nom}_1} - \frac{f^\textrm{carrier}_3 v_{31}}{f^\textrm{nom}_1 c} & 0 & 0 \\
0 \dots 0 & 0 & 0 & 0 & 0 & 1 & 0 & -1
\end{bmatrix}_{\hat{x}_k^-}, \nonumber
\end{eqnarray}
\end{widetext}
where we have omitted the step index $k$ in the components of the matrix. Notice that the dynamic matrix $F_k$ for this model
is a constant matrix, and most components of the measurement matrix $H'_k$ are constant. The nonlinearity of the Kalman filter
model has been significantly reduced. Also, the number of the variables in the state vector is 11, which is less than the number
of measured quantities 18. Except for the absolute frequency errors of the clocks, all other variables in the state vector are directly constrained
by the measurements, hence in principle they can be determined by the Kalman filter.

We design a hybrid-extended Kalman filter based on the model described above to process simulated LISA measurement data.
The estimation errors in the differential clock errors, arm lengths and the differential frequency errors are plotted in Fig.~\ref{fig:K_sin}.
With this phenomenological periodic model, the designed Kalman filter has successfully estimated these three kinds of interspacecraft
quantities. We also find that the designed Kalman filter model is not sensitive to the actual orbits of the LISA constellation.
With either angular velocities $\omega = 2\pi\,$rad/year or $\omega = 4\pi\,$rad/year and with either LISA orbits shown in
Fig.~\ref{fig:armlength} (a) or (b), the performance of the designed Kalman filter turns out to be similar in terms of estimation
errors.


\section{An effective system model}
\label{S:poly}

In this section, we wish to design another effective system model, whose errors can be accessed analytically. Compared to the periodic system model, this form
of system model is expected to be directly applicable to relativistic LISA orbits under consideration of all the planets in the solar system. The arm lengths of LISA are smooth, slowly varying and (nearly) periodic functions of time, no matter whether they are calculated in the non-relativistic or
relativistic framework. The smooth arm-length functions of time can be decomposed into harmonics
\begin{eqnarray}
L(t) = \bar{L} + \sum_{n=1}^{\infty} A_n \sin(n\omega t + \phi_n),
\end{eqnarray}
where $\bar{L}$ is the average arm length, and $A_1$ is the lowest order arm-length variation,
whose value is about $\bar{L}/100$ according to the orbit design of LISA.

The arm-length function can be expanded in polynomials around any time $t_0$
\begin{eqnarray}
L(t) &=& \bar{L} + \sum_{n=1}^{\infty} A_n \sin(n\omega t_0 + \phi_n) + \sum_{n=1}^{\infty} A_n \cos(n\omega t_0 + \phi_n) (n\omega \Delta t) \nonumber \\
&& \;\; - \frac{1}{2} \sum_{n=1}^{\infty} A_n \sin(n\omega t_0 + \phi_n) (n\omega \Delta t)^2 + O[\Delta t^3],
\end{eqnarray}
where $\Delta t \equiv t - t_0$. For elliptical Kepler orbits, $A_n$ decays exponentially with $n$. For relativistic orbits, $A_n$ also decays
much faster than linearly, which leads to $(n+1) A_{n+1} \ll n A_n $. For a short time $\Delta t = 1000\,$s, we estimate the contribution of each order
as follows
\begin{eqnarray}
A_1 \omega \Delta t &\sim& 10^4 \, \textrm{m},  \\
\frac{1}{2} A_1 (\omega \Delta t)^2 &\sim& 1 \, \textrm{m}, \\
\frac{1}{6} A_1 (\omega \Delta t)^3 &\sim& 10^{-4} \, \textrm{m}.
\end{eqnarray}
As a consequence, if we want to design a Kalman filter to process $1000\,$s LISA measurement data, the following
polynomial model characterize the LISA arm lengths to $0.1\,$mm accuracy.
\begin{eqnarray}
L(t) &=& \bar{L} + \sum_{n=1}^{\infty} A_n \sin(n\omega t_0 + \phi_n) + v \Delta t \nonumber \\
&+& \frac{1}{2} a \Delta t^2 + O[\Delta t^3],
\end{eqnarray}
where $v$ and $a$ are phenomenological variables.

According to the above phenomenological model, we define a 14-dimensional state vector in the Kalman filter as follow
\begin{eqnarray}
x=(L_{21}, v_{21}, a_{21}, L_{32}, v_{32}, a_{32}, L_{13}, v_{13}, a_{13}, \nonumber \\
\delta T_1 - \delta T_2, \delta T_2 - \delta T_3, \delta f_1, \delta f_2, \delta f_3)^T. \nonumber
\end{eqnarray}
The dynamic model for the arm-length phenomenological variables is simply as follows
\begin{eqnarray}
\frac{\mathrm{d}}{\mathrm{d}t}
\begin{bmatrix}
L \\
v \\
a
\end{bmatrix}
=
\begin{bmatrix}
0 & 1 & 0 \\
0 & 0 & 1 \\
0 & 0 & 0
\end{bmatrix}
\begin{bmatrix}
L \\
v \\
a
\end{bmatrix}
+ \mathrm{noise}.
\end{eqnarray}
The dynamics for the clock variables are the same as Eq.~\ref{eq:ClockEq}.

The measurement matrix $H_k = \left.\frac{\partial h_k}{\partial x}\right|_{\hat{x}_k^-}$ is 18-by-14 and a bit different from that in the
last section, we therefore again explicitly write its first three rows as below
\begin{widetext}
\begin{eqnarray}
\begin{bmatrix}
0  \dots  0 & 1 & 0 & 0 & c & c & 0 & 0 & 0 \\
0  \dots  0 & 0 & \frac{f_3^\textrm{carrier}}{c}\left( 1- \frac{\delta f_1}{f^\textrm{nom}_1}\right) & 0 & 0 & 0 & \frac{f^\textrm{carrier}_3-f^\textrm{carrier}_1}{f^\textrm{nom}_1} - \frac{f^\textrm{carrier}_3 v_{31}}{f^\textrm{nom}_1 c} & 0 & 0 \\
0  \dots  0 & 0 & 0 & 0 & 0 & 0 & 1 & 0 & -1
\end{bmatrix}_{\hat{x}_k^-}, \nonumber
\end{eqnarray}
\end{widetext}
where we have omitted the step index $k$ in the matrix components.

We design a hybrid-extended Kalman filter with the phenomenological polynomial system model described above to process
the simulated LISA measurement data. The results of the simulation are summarized in Fig.~\ref{fig:K_poly}. Comparing
these estimation errors with that given in the last section, we find that the overall performance of this phenomenological
polynomial model is slightly better. Since this model is directly applicable to relativistic cases, we expect it to outperform
the periodic model in general.

\begin{widetext}

\begin{figure}[htbp]
\centering
\subfloat[]{
\begin{minipage}[t]{0.33\textwidth}
\centering
\includegraphics[width=1.0\textwidth]{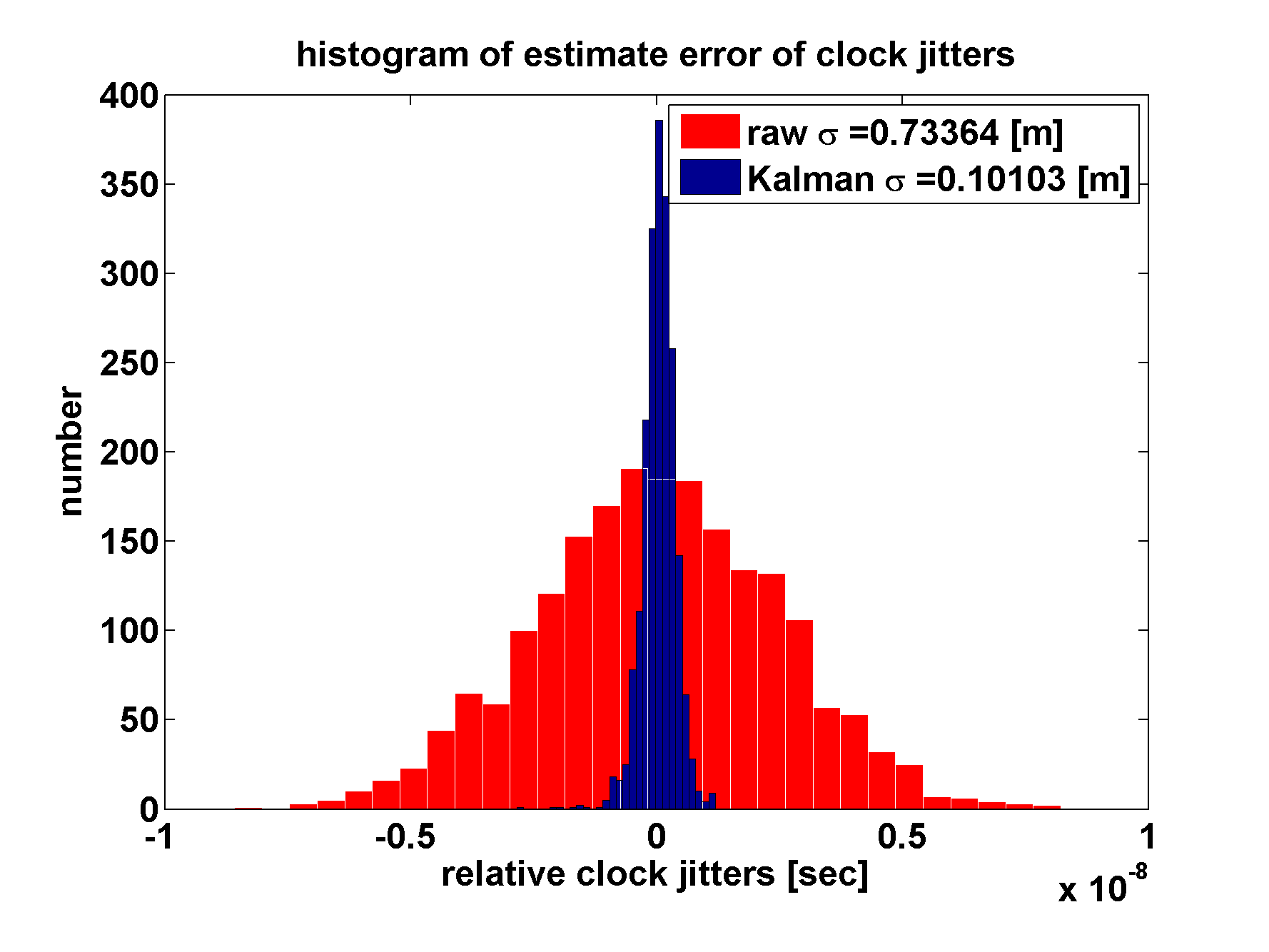}
\end{minipage}
}
\subfloat[]{
\begin{minipage}[t]{0.33\textwidth}
\centering
\includegraphics[width=1.0\textwidth]{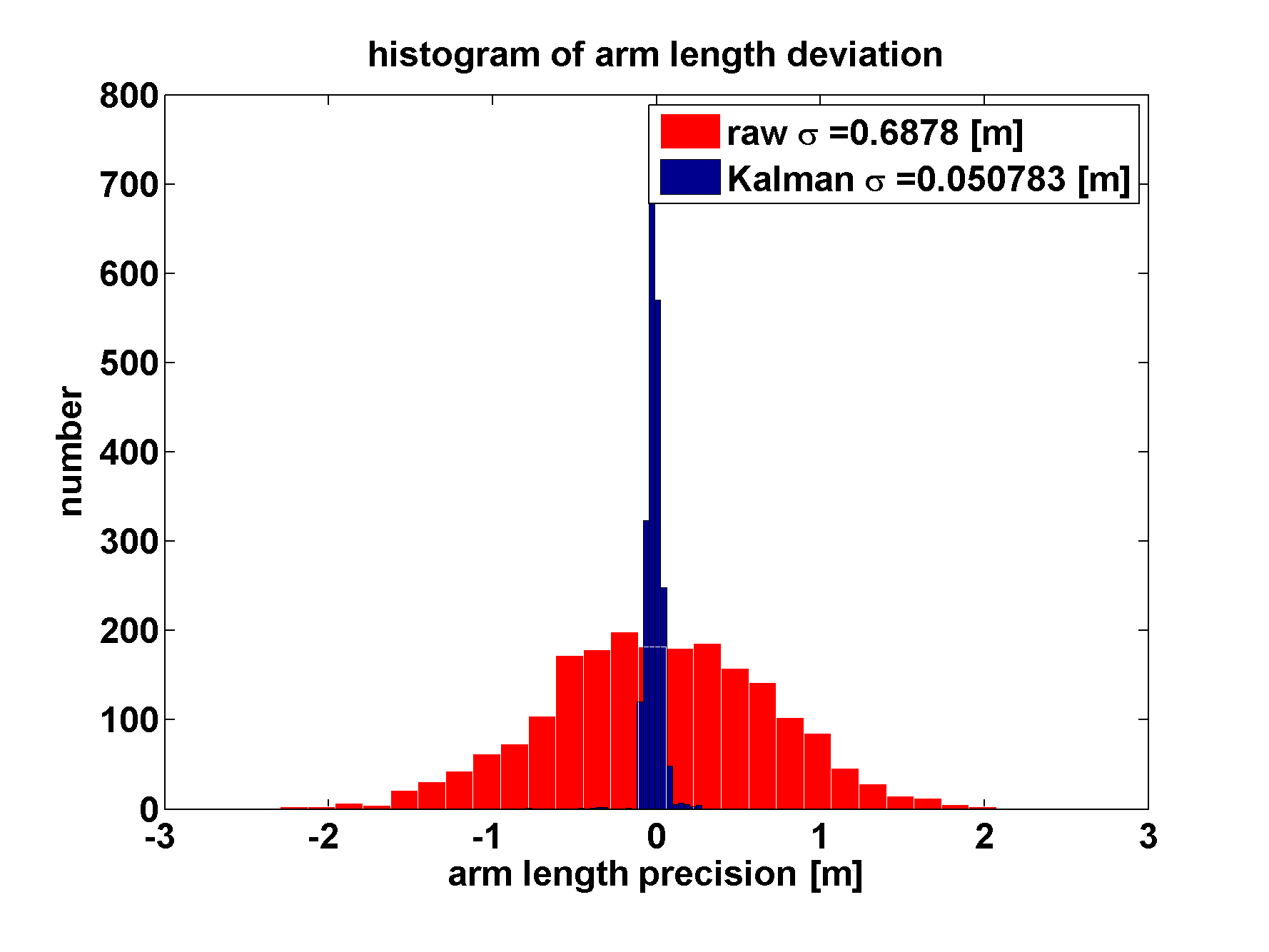}
\end{minipage}
}
\subfloat[]{
\begin{minipage}[t]{0.33\textwidth}
\centering
\includegraphics[width=1.0\textwidth]{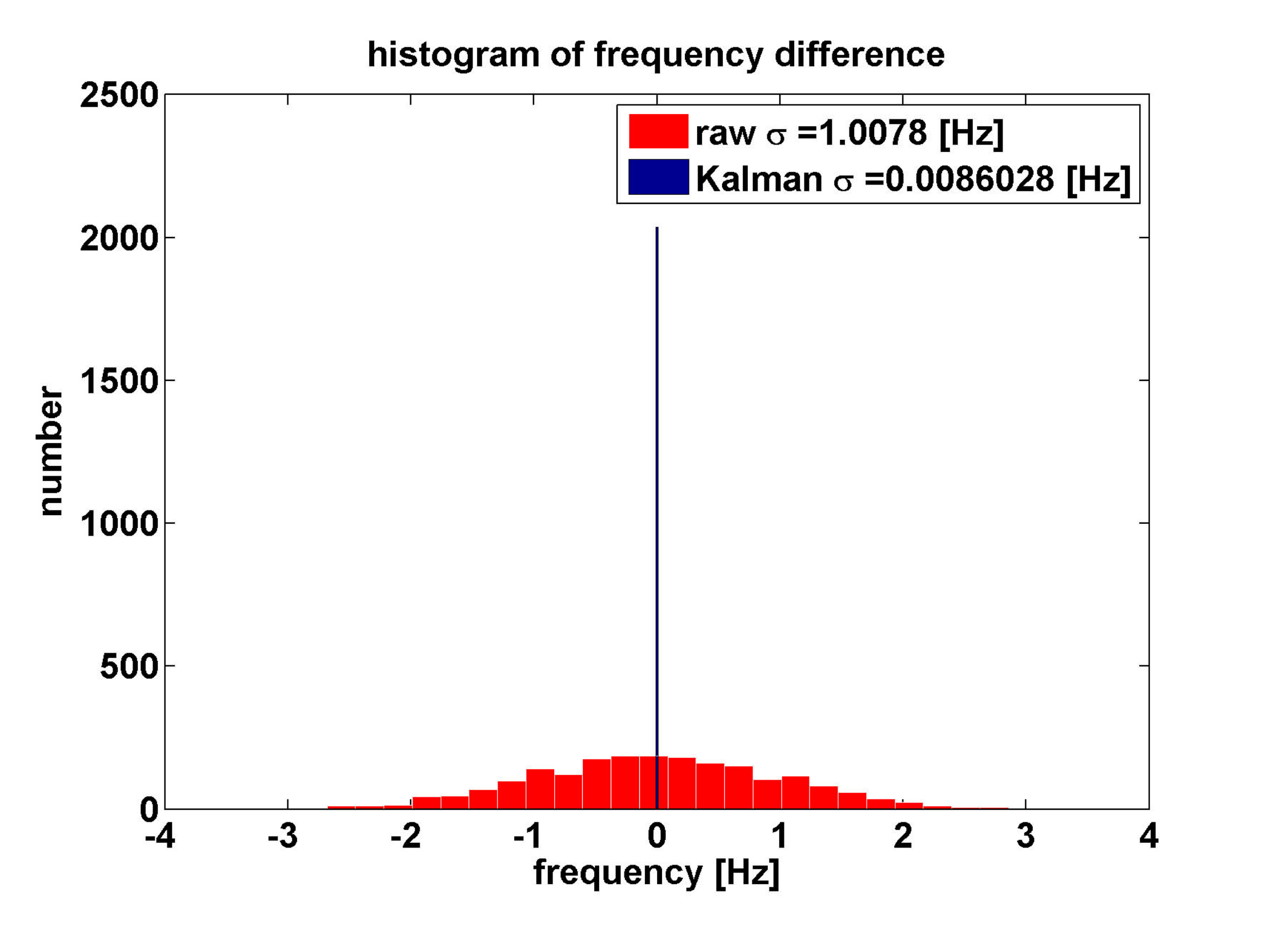}
\end{minipage}
}
\caption{ \label{fig:K_sin} Histograms of the estimation errors of a hybrid-extended Kalman filter with a periodic system model
in (a) differential clock errors (b) arm lengths and (c) differential frequency errors of the clocks for the laser link from S/C 2 to S/C 1. }
\end{figure}

\begin{figure}[htbp]
\centering
\subfloat[]{
\begin{minipage}[t]{0.33\textwidth}
\centering
\includegraphics[width=1.0\textwidth]{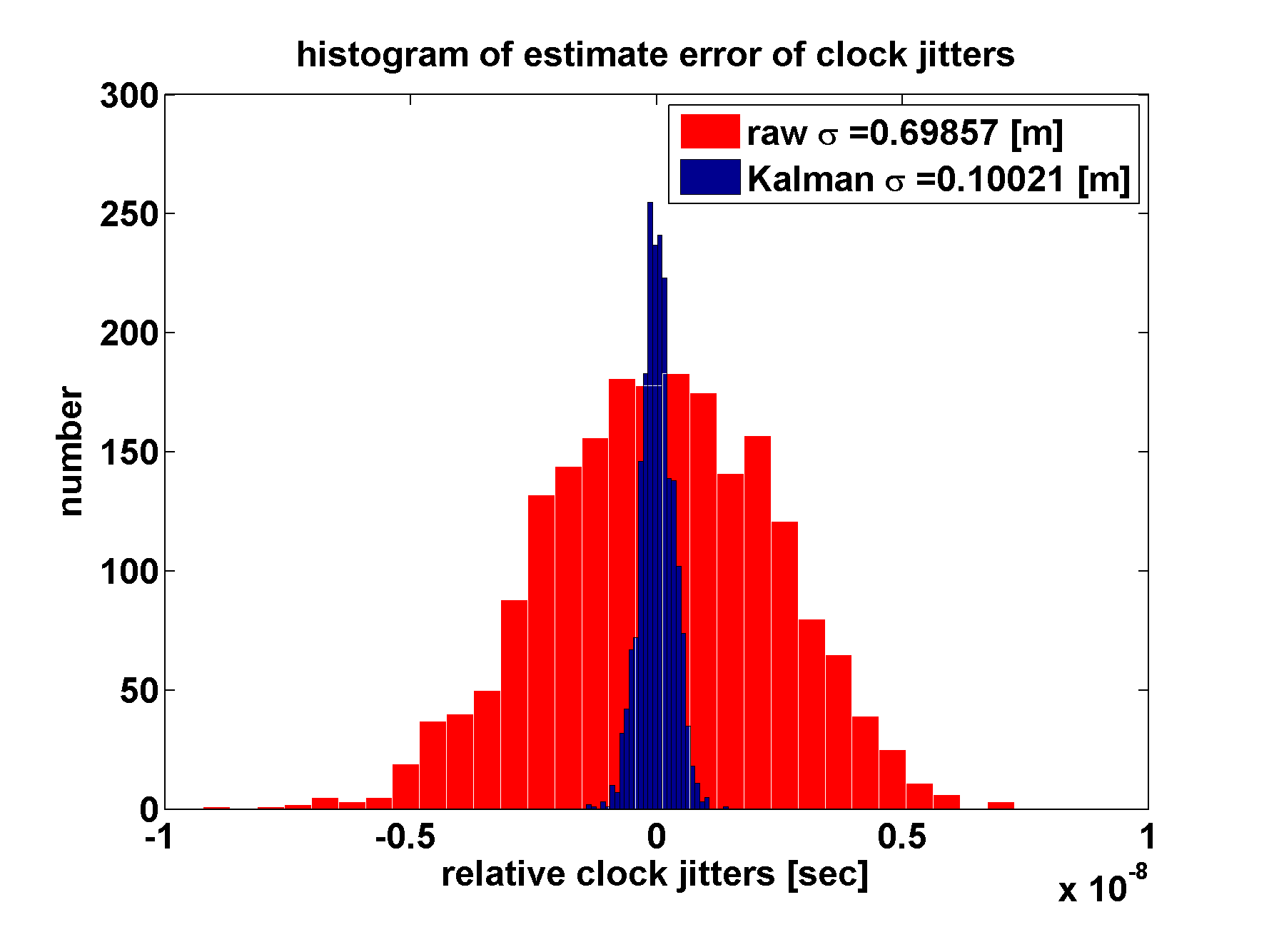}
\end{minipage}
}
\subfloat[]{
\begin{minipage}[t]{0.33\textwidth}
\centering
\includegraphics[width=1.0\textwidth]{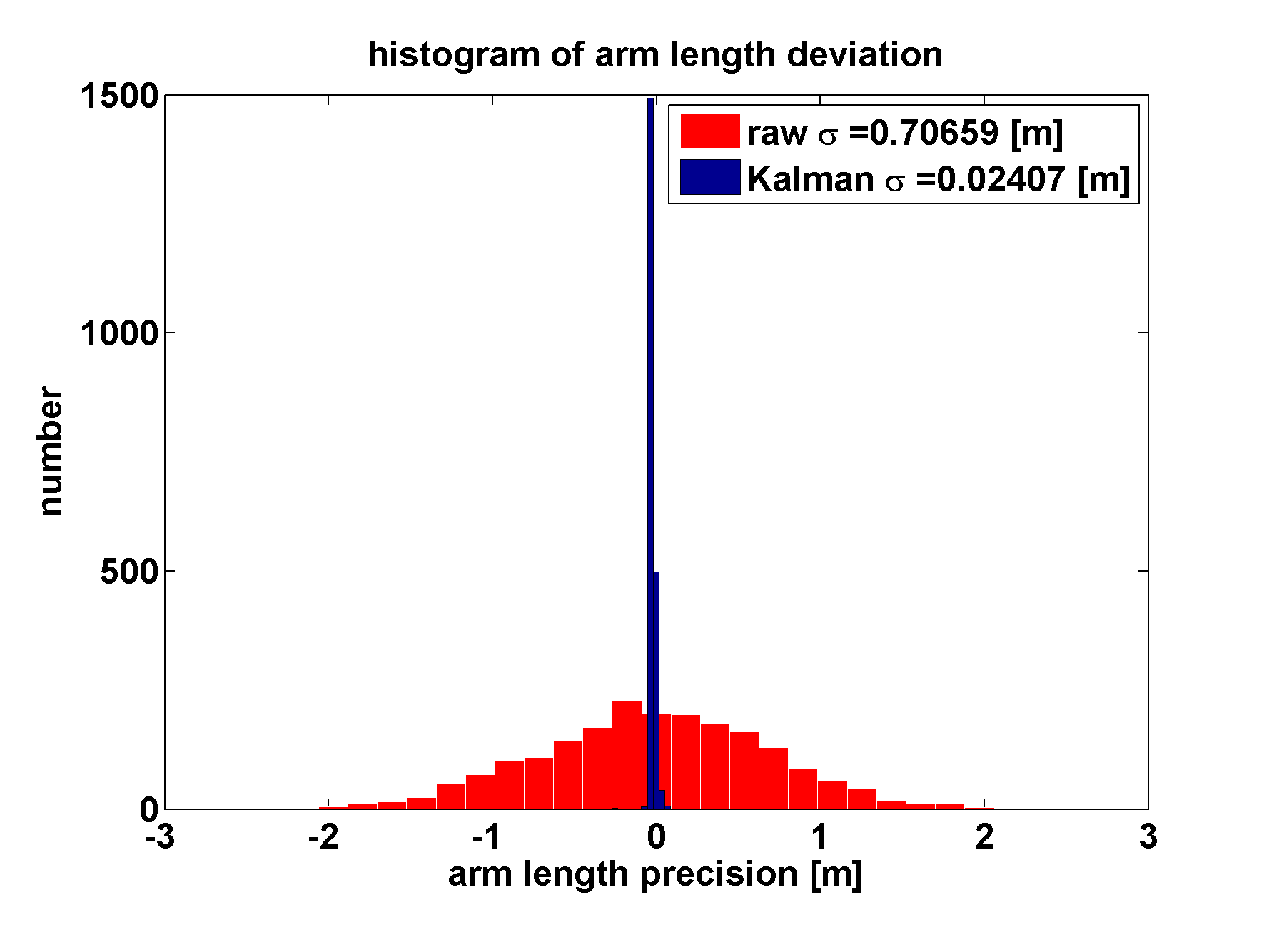}
\end{minipage}
}
\subfloat[]{
\begin{minipage}[t]{0.33\textwidth}
\centering
\includegraphics[width=1.0\textwidth]{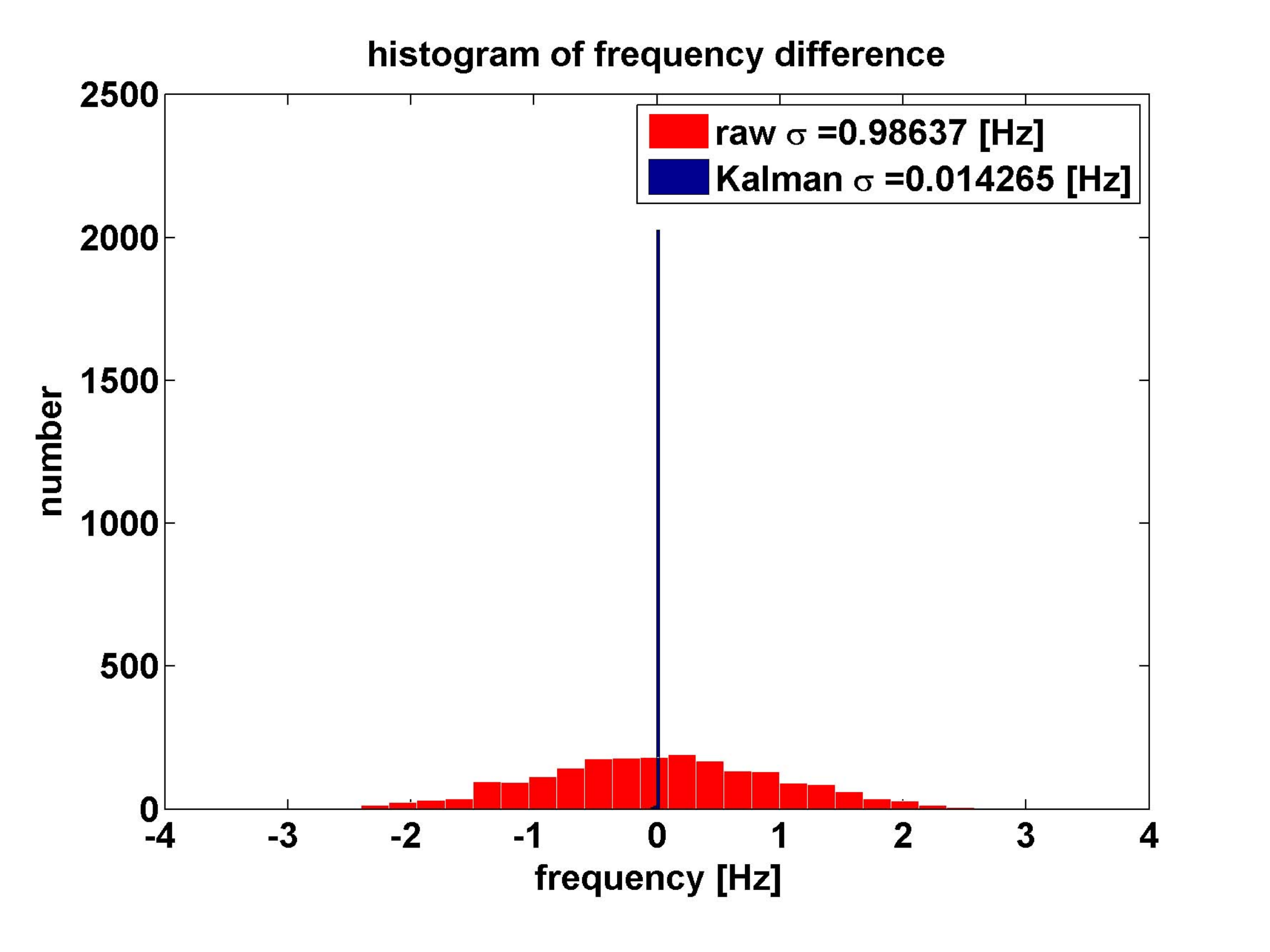}
\end{minipage}
}
\caption{ \label{fig:K_poly} Histograms of the estimation errors of a hybrid-extended Kalman filter with a phenomenological polynomial system model
in (a) differential clock errors (b) arm lengths and (c) differential frequency errors of the clocks for the laser link from S/C 2 to S/C 1. }
\end{figure}

\end{widetext}

\section{Summary}
\label{S:sum}

In this paper, we have designed and tested several alternative filtering dynamic models. The dynamic
models with 24-dimensional, 23-dimensional and 22-dimensional state vectors are comparable in reducing
the measurement noise. The greatly simplified Kalman filter model with only clock variables performs
a few times worse than the other filters as expected. All the filter models have successfully and
significantly reduced the measurement noise.

Most of the designed models use physical parameters, such as the positions and the
velocities of the S/C, in the state vector to characterize the motion of the LISA constellation.
The advantage of using these physical parameters is that they automatically fit into the
physical laws, hence we can directly use Newton's equations or relativistic equations as
the dynamic equations to evolve the state vector of the Kalman filter. However, LISA interspacecraft
measurements contain only relative quantities, such as the arm lengths, relative tangential velocities,
relative clock jitter and relative frequency jitter. Therefore, the number of variables in the state
vector is larger than the number of measurements, the physical variables in the state
vector cannot be fully determined, and the measurement equations are very nonlinear.
All of these add up complexities and numerical inaccuracies to the Kalman filter.

Two phenomenological models have been designed and simulated. These models are superior to the models
that directly use positions and velocities of the S/C, in the sense that they only use the observed
relative quantities in their state vector. In turn, the state vector is much smaller, and the propagation
matrices are much better conditioned. These models are also advantageous in reducing the nonlinearity
of the dynamic equations and readiness to be extended to more complicated cases, such as full relativistic
laser links.

\appendix

\section{A limitation on the common clock drift}

As mentioned before, the differential clock errors can be estimated very accurately, whereas the common mode of the clock errors cannot be determined.
This would result in errors in the time labels of the measurements. The errors in the time labels will introduce modulations to GW signals,
hence they may affect the detection and the parameter estimation of GW signals. In this appendix, we try to estimate this effect and
to set a limit on the permissible common clock drifts.

We define a quantity $\sigma_T(\tau)$ to characterize the timing stability
\begin{eqnarray}
\sigma_T (\tau) = \sqrt{\frac{1}{2\tau^2} \langle ( \delta T(k+1) - \delta T(k) )^2 \rangle },
\end{eqnarray}
where $\tau$ is the nominal time increment between the sample $k$ and the sample $k+1$.
Notice that this is different from the Allan variance \cite{Allan66}
\begin{eqnarray}
\sigma^2_A (\tau) &=& \frac{1}{2} \left\langle \left( \frac{\delta f(k+1)}{f^\textrm{nom}} - \frac{\delta f(k)}{f^\textrm{nom}} \right)^2\right\rangle \nonumber \\
&=& \frac{1}{2\tau^2} \langle ( \delta T(k+2) - 2 \delta T(k+1) + \delta T(k) )^2 \rangle ,  \nonumber \\
\end{eqnarray}
which characterizes the frequency stability of the clock.

For a GW signal with frequency $f_\textrm{GW}$ and a total observation time $T_\textrm{obs}$, if the mismatch caused by
the common mode of the clock errors is less than $\epsilon$ cycle, the effect on physical data analysis is negligible. Thus, we have
a limitation on the common clock drift
\begin{eqnarray}
\sigma_T(\tau)f_\textrm{GW}T_\textrm{obs}<\epsilon.
\end{eqnarray}
For physical data analysis concern, a sampling time of $\tau=1\,$s will suffice. By considering the worst
scenario, we take $f_\textrm{GW}=0.1\,$Hz, $T_\textrm{obs}=10^8\,$s and $\epsilon=0.1$, hence the timing
stability should satisfy $\sigma_{T}(1)<10^{-8}$.

\begin{acknowledgements}
The authors are partially supported by DFG Grant No. SFB/TR 7 Gravitational Wave
Astronomy and the DLR (Deutsches Zentrum f\"ur Luft- und
Raumfahrt). The authors would like to thank the German Research Foundation for funding
the Cluster of Excellence QUEST-Center for Quantum Engineering and
Space-Time Research.
\end{acknowledgements}

\end{document}